\documentclass[]{aa}
\usepackage{graphicx}

\usepackage[colorlinks=true,linkcolor=red,urlcolor=red,citecolor=blue]{hyperref}

\usepackage{txfonts}

\newcommand{\myemail}{antonello.calabro@inaf.it}

\usepackage{multirow,array}

\usepackage{tikz}

\definecolor{lime}{HTML}{A6CE39}
\DeclareRobustCommand{\orcidicon}{%
    \begin{tikzpicture}
    \draw[lime, fill=lime] (0,0) 
    circle [radius=0.16] 
    node[white] {{\fontfamily{qag}\selectfont \tiny ID}};
    \draw[white, fill=white] (-0.0625,0.095) 
    circle [radius=0.007];
    \end{tikzpicture}
    \hspace{-2mm}
}

\foreach \x in {A, ..., Z}{%
    \expandafter\xdef\csname orcid\x\endcsname{\noexpand\href{https://orcid.org/\csname orcidauthor\x\endcsname}{\noexpand\orcidicon}}
}

\usepackage[rightcaption]{sidecap}

\begin{document}

\title{Investigating the role of mergers in galaxy assembly in the early Universe ($z>5$)}  

\author{
A.Calabr{\`o}\inst{1}   
\and L.Pentericci\inst{1}
\and M.Llerena\inst{1}
\and S.Rossi\inst{2,1}
\and L.Napolitano\inst{1}
\and D.Bevacqua\inst{1}
\and M.Giavalisco\inst{3}
\and R.Somerville\inst{4}
\and G.Gandolfi\inst{1}
\and E.Daddi\inst{5}
\and M.Dickinson\inst{6}
\and S.Finkelstein\inst{7}
\and A.Fontana\inst{1}
\and M.Hirschmann\inst{8}
\and J.S.Kartaltepe\inst{9}
\and D.Kocevski\inst{10}
\and A.M.Koekemoer\inst{11}
\and H.Leung\inst{12}
\and R.A.Lucas\inst{11}
\and A.Taylor\inst{7}
\and R.Tripodi\inst{1}
\and X.Wang\inst{13,14,15}
\and L.~Y.~A.~Yung\inst{11}
}

\institute{INAF - Osservatorio Astronomico di Roma, via Frascati 33, 00078, Monte Porzio Catone, Italy (\myemail) % 1
\and  Dipartimento di Fisica, Sapienza, Università di Roma, Piazzale Aldo Moro 5, 00185, Roma, Italy %2
\and University of Massachusetts Amherst, 710 North Pleasant Street, Amherst, MA 01003-9305, USA % 3
\and Center for Computational Astrophysics, Flatiron Institute, 162 5th Avenue, New York, NY 10010, USA % 4
\and Universit{\'e} Paris-Saclay, Universit{\'e} Paris Cit{\'e}, CEA, CNRS, AIM, 91191, Gif-sur-Yvette, France %5
\and NSF's National Optical-Infrared Astronomy Research Laboratory, 950 N. Cherry Ave., Tucson, AZ 85719, USA %6
\and The University of Texas at Austin, 2515 Speedway Blvd Stop C1400, Austin, TX 78712, USA % 7
\and Institute of Physics, Laboratory of Galaxy Evolution, Ecole Polytechnique F{\'e}d{\'e}rale de Lausanne (EPFL), Observatoire de Sauverny, 1290 Versoix, Switzerland %8
\and Laboratory for Multiwavelength Astrophysics, School of Physics and Astronomy, Rochester Institute of Technology, 84 Lomb Memorial Drive, Rochester, NY 14623, USA %9
\and Department of Physics and Astronomy, Colby College, Waterville, ME 04901, USA %10
\and Space Telescope Science Institute, 3700 San Martin Drive, Baltimore, MD 21218, USA %11
\and SUPA, Institute for Astronomy, University of Edinburgh, Royal Observatory, Edinburgh EH9 3HJ, UK % 12
\and School of Astronomy and Space Science, University of Chinese Academy of Sciences (UCAS), Beijing 100049, China %13
\and National Astronomical Observatories, Chinese Academy of Sciences, Beijing 100101, China %14
\and Institute for Frontiers in Astronomy and Astrophysics, Beijing Normal University, Beijing 102206, China %15
}
\date{Submitted A\&A}

\abstract 
{
Galaxy mergers play a crucial role in shaping the morphology, the star formation, and the mass growth of galaxies across cosmic time. While mergers have been extensively investigated in the local Universe, the evolution of their frequency and physical properties in the early Universe has yet to be fully understood. In this work, we investigate the role of mergers in a large spectroscopic sample of 1233 galaxies in the range $5<z<14$ with good detection (S/N$_{\rm pixel} > 3$) in the JWST/F444W band, covering six different extragalactic fields (COSMOS, UDS, GOODS-South, GOODS-North, EGS, and Abell2744). %to ensure reliable morphological measurements. 
We identify mergers from rest-frame optical disturbances in F444W, using a combination of Gini, M$_{20}$, and Asymmetry morphological parameters, which trace more advanced mergers (including also minor mergers) with shorter observability timescales compared to the close photometric pair selection. 
The observed morphological merger fraction f$_m$ does not strongly evolve with redshift from $z=0$ to $z \sim 8$. We find an average merger fraction of f$_m$ $\sim 5 \%$ for our golden merger condition (Gini$+0.14\times$M$_{20}>0.33$, A$>0.35$) tracing major mergers with $\tau_{\rm obs,golden}$ $30$-$10$ Myr at $z=$ $5$-$10$. This fraction increases to $\sim 8\%$ for asymmetry mergers ($A > 0.35$) owing a longer observability timescale, and reaches $\sim 13\%$ when considering Gini-M$_{20}$ mergers (Gini$+0.14\times$M$_{20}>0.33$), which trace major+minor mergers. 
After accounting for the different merger observability timescales, and considering the shorter merger duration in the early Universe, we find that the merger rate is strongly increasing from $z=1$ to $z=7$ by more than one order of magnitude, averaging $\sim 2$ merger/galaxy/Gyr at redshifts $5<z<10$ for major mergers (in agreement with photometric pair studies), and a factor of $\sim 3$ higher for minor+major mergers. 
We finally perform SED modeling using available HST+JWST photometry to infer stellar masses and star-formation rates (SFRs) averaged over multiple timescales, using a non parametric star-formation history. 
We find that mergers at this cosmic epoch have a significant impact (although significantly lower than at $z=1$) on the SFR of galaxies, with an average enhancement by a factor of $\sim 1.8$  compared to isolated galaxies in the same mass and redshift range. This enhancement is higher in the last $10$ Myr and for shorter merger timescale indicators, suggesting that mergers trigger star-formation through short-lived powerful bursty episodes. Despite this, mergers contribute only by $5 \%$-$10 \%$ to the mass build-up of galaxies in the redshift range explored, suggesting that in-situ star-formation is the dominant process in galaxy growth at all cosmic epochs.}


\titlerunning{\footnotesize The role of mergers in galaxy assembly from redshift $5$ to $12$}
\authorrunning{A.Calabr\`o et al.}
 \maketitle
\section{Introduction }\label{sec:introduction}

Galaxy mergers play a fundamental role in the evolution of galaxies over cosmic time. In $\Lambda$-CDM cosmology, they are a key mechanism for hierarchical structure formation, through which small dark matter halos and their host galaxies merge together to form larger halos, larger galaxy structures and more massive galaxies \citep{kormendy04,hopkins08}. Mergers also drive profound modifications of galaxies' morphologies, as they are key channels for the conversion of spiral galaxies into elliptical and lenticular ones \citep{toomre72,barnes92,naab14}. In the low redshift Universe, they also impact the star-formation rate (SFR) of the galaxies. In the SDSS survey, \citet{patton13} found a SFR enhancement of a factor of $\sim2.5$ in mergers compared to isolated systems at $z \simeq 0$. Using HST images, a similar result was found in close pairs with projected separations of $< 5$kpc by \citet{shah22} at $z=1$, while at higher redshifts ($1 < z < 3$) the SFR enhancement decreases to a factor of $\times 1.5$. In contrast, \citet{pearson19b} do not find any significant enhancement of the SFR in mergers compared to isolated galaxies in a sample of 200k galaxies at $0<z<4$. 

Recently, JWST has revolutionized the study of galaxy morphology and mergers in the early Universe ($z>5$) by providing near-infrared images with unprecedented depth and resolution in different extra-galactic fields (see details on the first large imaging campaigns in \Citealt{treu22}, \Citealt{casey23}, \Citealt{bezanson24}, \Citealt{finkelstein25}, \Citealt{eisenstein26}), revealing details in the rest-frame UV and optical (e.g., faint interacting features, double nuclei) that were not accessible with other space-based and ground-based facilities. This has allowed the astronomical community to study the morphology of early galaxies \citep[e.g. ,][]{treu23,kartaltepe23} and identify sizable samples of mergers during the reionization epoch (EoR), thus exploring the relevance of mergers and their impact on the SFR during the first Billion year of cosmic time. 

Different approaches can be used to identify mergers. The first, simpler approach relies on selecting galaxy pairs that are close on the sky (up to a maximum separation distance) and with consistent photometric redshifts (e.g., \Citealt{mundy17}, \Citealt{duncan19}). Using this approach on JWST observations, several works have found that the merger rate at $z>5$ is orders of magnitude higher than in the local Universe \citet{duan25,puskas25_mergers}. According to \citet{duan25}, mergers also play an important role in the galaxy mass accretion at that epoch, contributing $\sim 50 \%$ to the mass build-up of galaxies, comparable to the mass accretion from star formation. However, this result is not confirmed by other studies, with \citet{puskas25_mergers} finding a lower merger contribution of $\sim10\%$.

The impact of mergers on the SFR at the EoR is also a hotly debated issue. \citet{puskas25_SFR}, based on $\sim 2000$ galaxies in the JADES survey at redshifts $3$-$9$, find that photometric pairs with $<20$ kpc projected separations have a specific SFR (sSFR) enhancement by a factor of $1.12 \pm 0.05$ compared to isolated galaxies at the same stellar mass (M$_\ast$) and redshift. Although this effect is smaller at high-z compared to the local Universe, mergers still have a significant impact on the SFR at $z>3$. In contrast, \citet{duan26} found no statistical enhancement in sSFR for close pairs at any projected separation at $z>5$. 

Despite the relative simplicity of the close photometric pair approach, it presents several limitations. First, it mainly traces the early phases of galaxy interactions rather than advanced mergers, hence providing a limited and incomplete picture of the merger phenomenon. Since it does not rely on the detection of galaxy distortions or asymmetries, the true interacting nature of the system and its final fate crucially depend on the accuracy of photometric redshifts and on the relative velocity of member galaxies.  

Alternative approaches are able to select mergers in more advanced phases of the interaction. 
These include the visual inspection of the galaxy morphology, which has been succesfully adopted by \citet{kartaltepe15} for morphological classification in the CANDELS fields with HST images, and more recently applied also on JWST observations of galaxies in the CEERS survey up to $z\sim9$ \citep{kartaltepe23}. However, the main limitations of this approach are reproducibility and feasibility for large samples of galaxies. 
In order to solve these limitations, machine learning techniques have also been adopted in the last decade for merger selection \citep[e.g.,][]{huertas-company15,huertas-company18,pearson19a,margalef-bentabol24,ferreira20}. However, the final classification crucially depends on the dataset that is used for training the algorithm. Even though hydrodynamical simulations can also be adopted to define the ground truth on the merger nature of a system, projects based on this approach have shown contrasting results, with accuracies of the final merger selection ranging from $40 \%$ to $\sim 75 \%$ \citep{guzman-ortega23,rose23}. 

Another approach that is feasible for large galaxy samples is based on the measurement of some fundamental morphological parameters, namely the Asymmetry (A), Gini, and M$_{20}$ parameters, which were introduced by \citet{conselice03} and \citet{lotz04} to identify mergers from HST images of galaxies at $z < 3$. These parameters are highly sensitive to the presence of a disturbed and asymmetric structure, features due to interactions such as connecting bridges and tidal tails, shells, and double nuclei embedded in a common, more diffused gaseous and stellar envelopes. They are thus well suited to identify evolved, pre-coalescence, merger stages, complementary to the close pair selection. Recently, they have been successfully adopted in several works to identify morphological disturbances in galaxies from JWST images \citep[e.g.,][]{liang24,polletta24,westcott25}. 
This quantitative morphological analysis has also been adopted by Dalmasso et al. 2024 to study the fraction and properties of mergers in a sample of $675$ photometrically selected galaxies at $4<z<9$ in the background of the Abell2744 cluster field. They find no significant redshift dependence of the merger fraction in the redshift range analyzed, and no significant difference in the sSFR between merging and non-merging galaxies. 
However, their sample could still be contaminated by low-redshift interlopers, which could potentially have a significant impact both on the merger rate and on their average properties, especially at the highest redshifts where the statistics are very poor. 

In this paper, we expand on the study of \citet{dalmasso24}, revisiting the role of mergers in galaxy mass assembly and determining to what extent mergers affect star formation during the EoR. Compared to this previous work, we complement the non parametric morphology approach with a visual inspection to increase the purity of the merger selection. We also improve the statistics by considering all the CANDELS fields, namely GOODS-North, GOODS-South, UDS, EGS, and COSMOS. Moreover, given the larger number of spectroscopic confirmations and observations that are now publicly available from JWST, we select galaxies spectroscopically in order to avoid contamination from low-redshift interlopers, while retaining a good sample of statistics for a robust analysis. Despite this strictier requirement, we extend the galaxy sample of \citet{dalmasso24} by a factor of $\sim 2$. The availability of spectroscopic data also allows us to study the emission line properties and ionizing source of merging galaxies, which will be investigated in a subsequent work. 

Our paper is organised as follows. In Section ~2, we present our spectroscopic sample selection and our methodology to assess galaxy morphologies. We also describe the SED fitting procedure through which we derive fundamental galaxy parameters, including stellar masses and specific star-formation rates (SFRs). In Section ~3, we present our results on the merger fraction evolution at $z > 5$ and on the SFR enhancement in mergers versus isolated galaxies. In Section ~4, we discuss our results and compare them to predictions of semi-analytic models and simulations. Finally, we present a summary and conclusions in Section ~5. 

Throughout the paper, we adopt the AB magnitude system \citet{oke83}, we assume a flat $\Lambda$CDM cosmology with $\Omega_{\rm m} = 0.3$, $\Omega_\Lambda = 0.7$, and $H_{0}=70$ $\rm km\ s^{-1}Mpc^{-1}$, and a \citet{chabrier03} initial mass function (IMF). We use the Astropy package \citep{astropy22}, the photutils package \citep{bradley23} \footnote{\url{https://photutils.readthedocs.io/en/stable/citation.html}}, and the statmorph python package \citep{rodriguez-gomez19a,rodriguez-gomez19b} for the calculation of morphological parameters.

\section{Methodology }\label{methodology}

\subsection{Spectroscopic sample selection}\label{sec:spectroscopic_sample_selection} 

\begin{figure}[t!]
    \centering
    \includegraphics[angle=0,width=1\linewidth,trim={0.cm 0.4cm 0.cm 0.cm},clip]{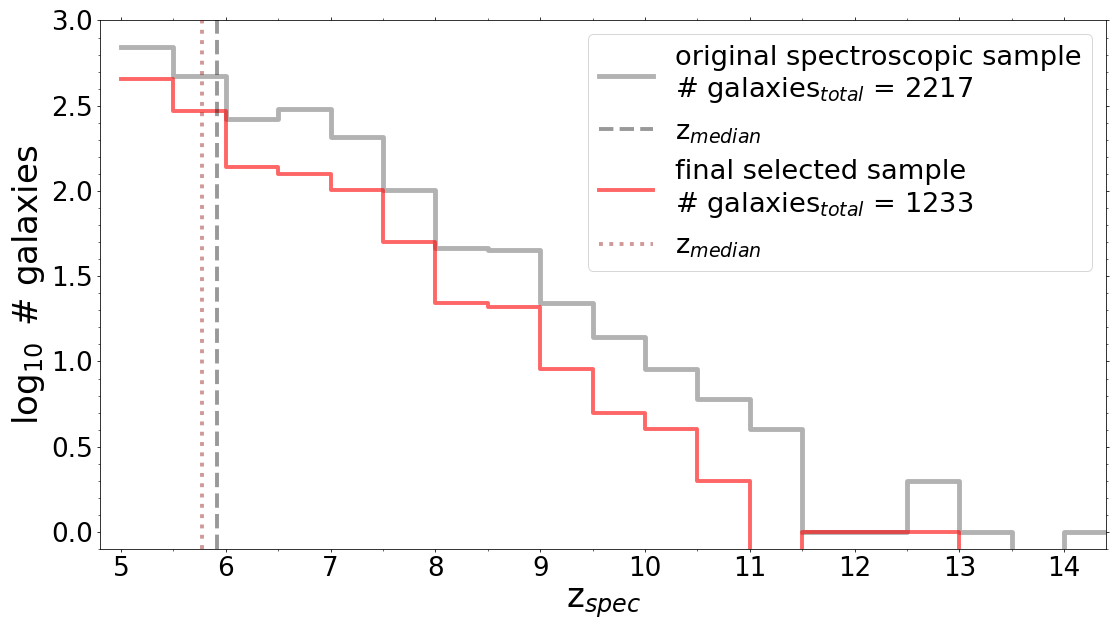}
    \vspace{-0.15cm}
    \includegraphics[angle=0,width=1\linewidth,trim={0.cm 0.cm 0.2cm 0.4cm},clip]{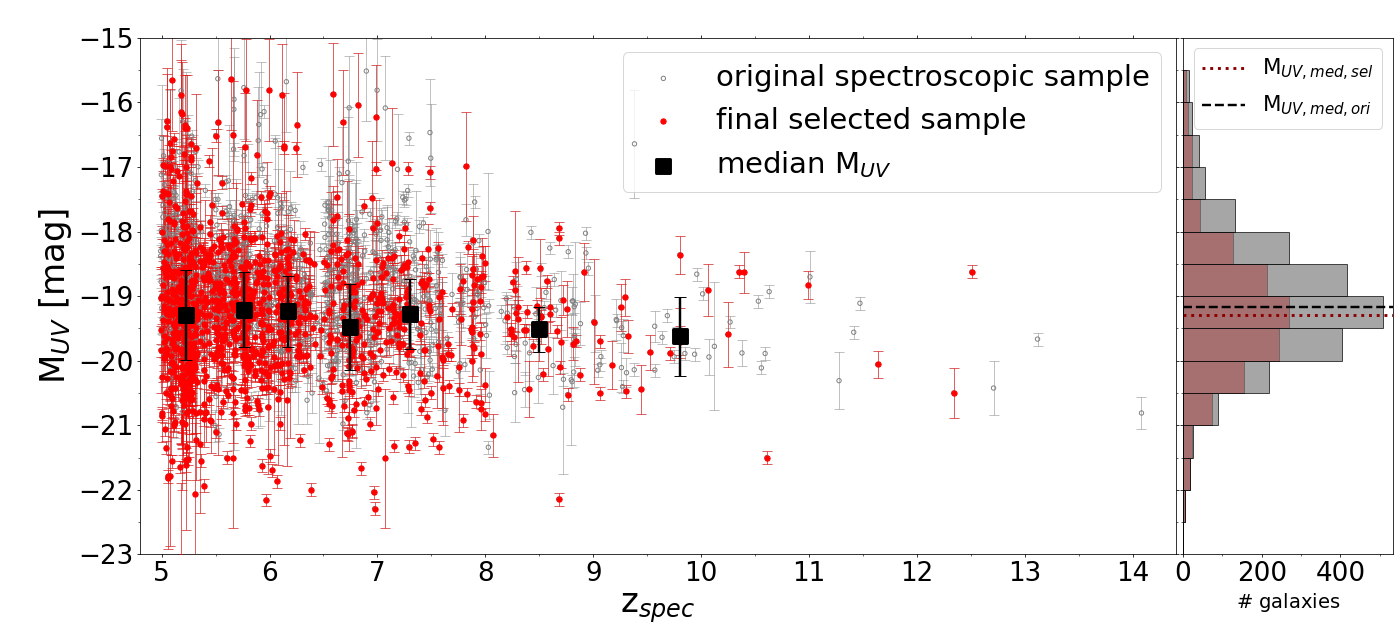}
    \vspace{-0.4cm}
    \caption{\textit{Top:} Histogram of the redshift distribution of the original spectroscopic sample selected from DJA with $z_{spec} > 5$ (gray line) and of the final selected sample after applying the S/N$_{\rm pixel}$ cut (red line). \textit{Bottom:} Scatter plot highlighting the M$_{\rm UV}$ distribution as a function of redshift for the original spectroscopic sample and the final sample (gray and red circles, respectively). The median values of M$_{\rm UV}$ in increasing redshift bins for the selected sample are highlighted with big dark red empty squares. The histogram of M$_{\rm UV}$ for the original and selected samples are also shown on the y right axis (respectively in gray and red), with the median M$_{\rm UV}$ ($=-19.3$ in both cases) indicated by a horizontal dashed and dotted line, respectively. 
    }\label{fig:histograms_selected_sample}
    \vspace{-0.3cm}
\end{figure}

The galaxy sample analyzed in this work is drawn from the DAWN JWST Archive (DJA \footnote{\url{https://dawn-cph.github.io/dja/blog/2025/05/01/nirspec-merged-table-v4/} }), which is an online public repository including the most recent JWST data products. In particular, we considered the latest version (v4.4) of the DJA spectroscopic catalog (DJA-Spec, \Citealt{valentino25}, \footnote{\url{doi:10.5281/zenodo.15472354}}), which includes spectroscopic observations and best-fit spectroscopic redshifts obtained from the Early Release Science (ERS) surveys to the most recent surveys performed in Cycle 3 with public data, such as CAPERS. 
We included the data taken in six different fields, including EGS, GOODS-SOUTH, GOODS-NORTH, UDS, and COSMOS, which were originally part of the CANDELS fields \citep{grogin11,koekemoer11}, with the addition of the field covering the background of the galaxy cluster Abell2744, which was part of the Hubble Frontier Fields \citep{coe15,lotz17}, and the adjacent GLASS field \citep{treu22}. The $5 \sigma$ depth of the imaging surveys (using a $0.1 \arcsec /$ radius circular aperture) ranges between m$_{AB}=$ $28$ and $30$ in F444W, specifically $\sim 28$ in COSMOS and UDS \citep{shuntov25,donnan24}, $\sim 29$ in EGS \citep{finkelstein23}, and $\sim 30$ in the GOODS fields and GLASS \citep{harvey25,treu22}.
The spectra were taken by several surveys during the first three cycles, the largest of which include CEERS (ERS-1345; \Citealt{finkelstein23,finkelstein25}), GLASS (ERS-1324 and DDT-2756; \Citealt{treu22,mascia24}), JADES (GTO-1180, 1210, GO-3215; \Citealt{bunker24,eisenstein25}), UNCOVER (GO-2561; \Citealt{bezanson24}), RUBIES (GO-4233, \Citealt{degraaff25}), and CAPERS (GO-6368, P.I. M.Dickinson). In these fields, deep, multiwavelength JWST-NIRCam imaging is also available from the same surveys. 

The spectra assembled in DJA were reduced with the GRIZLI \citep{brammer23a} and MSAExp \citep{brammer23b,heintz25} software, following a uniform spectroscopic reduction procedure that is described in detail in \citet{heintz23} and \citet{degraaff24}. The v4 release delivers to the community merged FITS files including all the 1D `optimally' extracted spectra available from different NIRSpec-MSA observations with the grating and prism configuration. 
We checked the photometric consistency of the NIRSpec-PRISM spectra, using the photometry available in the ASTRODEEP-JWST multi-band photometric catalog of \citet{merlin24}, correcting for residual flux calibration inconsistency (e.g., due to inaccurate aperture correction, instrumental instability). 
However, as already outlined in \citet{heintz25}, the 1D spectra retrieved from DJA already show an overall agreement within $10$-$15 \%$ with the photometric data. 

As a preliminary selection, we considered galaxies in the DJA-Spec catalog in the six fields listed above and with a robust (quality flag $=3$) spectroscopic redshift $\geq 5$, also confirmed through a visual inspection.  This redshift limit was chosen because it represents approximately the redshift limit up to which mergers and merger fractions were estimated by \citet{kartaltepe07} and \citet{duncan19} from previous HST surveys in the CANDELS fields. This limit also allows us to compare our findings to similar JWST studies performed at the same redshifts, and based on the photometric close pair method \citep[e.g.,][]{puskas25_mergers,duan25}. With these criteria, we obtained an initial sample of $2217$ galaxies. The redshift distribution of this sample is presented in Fig. \ref{fig:histograms_selected_sample}-\textit{top} (gray line), and shows that the number of sources decreases exponentially from $z=5$ to $z=14$, with a median redshift $z_{\rm med} \simeq 6$. 
Lastly, we used the galaxy spectra to derive the UV magnitudes M$_{ \rm UV}$ at $1500$ \AA\ rest-frame, using the methodology outlined in \citet{dottorini25}. We find M$_{ \rm UV}$ ranging between $-22.5$ and $-15.5$ (Fig. \ref{fig:histograms_selected_sample}-\textit{bottom}), with median M$_{ \rm UV}$ values that are not significantly evolving with redshift. 

\subsection{Derivation of morphological parameters from NIRCam images}\label{sec:morphological_parameters_derivation} 

We derive the morphological parameters of the galaxies from $2.5 \arcsec \times 2.5 \arcsec $ cutouts \footnote{We download the cutouts using the grizli cutout service: \url{https://grizli-cutout.herokuapp.com/}} in the NIRCam F444W band. The main motivation of this choice is that F444W traces the optical rest-frame for almost the entire redshift range spanned by our sample, thus the galaxy structure and merger identification are more directly comparable to lower redshift studies (at $z<3$) based on the HST/F160W or F814W band. We note, however, that the spatial resolution of JWST at $4.4 \mu m$ is better than the one of HST at $1.6 \mu m$ ($0.166 \arcsec$ vs $0.182\arcsec$, \Citealt{merlin24}), allowing us to detect finer and smaller galaxy structures. Bluer NIRCam bands, despite the higher resolution, mainly trace the rest-frame UV emission in our redshift range, which might be more affected by inhomogeneous dust distribution and might not trace real interacting features or merger driven asymmetric structure \citep{calabro19,cibinel19}. 

The morphological parameters selected for the merger identification are the Gini, the M$_{20}$, and the asymmetry parameters. We calculate them through \textit{statmorph}, a python package for the quantitative assessment of galaxy structure, and which is mostly based on the code written in IDL by \Citealt{lotz04} and \Citealt{lotz08_simul}. The definitions of these parameters are based on the original works of \citet{conselice03} and \citet{lotz04}, and recently reported in our previous studies \citep{treu23,dalmasso24}. However, we report them in the Appendix \ref{appendix1:definition_of_morphological_parameters} for completeness. For each cutout, we also derive the segmentation map of the sources following the procedure described in detail in \citet{treu23}. In brief, we use the photutils astropy package \citep{bradley23} to identify the sources as regions of at least five connected pixels with a flux of $2\sigma$ above the background in a smoothed version of the original cutout (using a $6 \times 6$) uniform filter.

An important step of the morphological analysis is the visual inspection of the cutout and removal of low redshift interlopers that wrongly fall in the same segmentation map of the main central galaxy. This is done through the comparison of the original segmentation maps derived with our procedure to the photometric catalog of \citet{merlin24}. In case of a system with two clearly separated galaxies belonging to the same segmentation map, we checked whether the photometric redshift (or the spectroscopic one, if available) of the offset component is consistent with the spectroscopic redshift of the central galaxy. 
We remove the secondary galaxy from the segmentation map through a deblending procedure in case of inconsistent redshifts, that is, $|z_{\rm phot,secondary} - z_{\rm spec,central}| > 0.15 \times (1+z_{\rm spec,central})$, as this is the condition for photo-z outliers in \citet{merlin24}. In these cases, we then rerun the morphological parameters routine.

Finally, for each galaxy, we also estimate the average S/N per pixel (S/N$_{\rm pixel}$) within its segmentation map, which we will use to assess the accuracy and reliability of the morphology calculations, as we explain in the following section.

\subsection{Reliability of morphological parameters}\label{sec:reliability}

When observing galaxies at increasing redshifts, we should consider that their structural appearance might be significantly altered because of surface brightness dimming, which roughly scales with redshift as ($1 + z$)$^{-4}$ \citep[e.g.,][]{whitney20}. This could potentially bias the estimation of morphological parameters. 
We note that in those extreme cases when the galaxies have a very low surface brightness and the signal-to-noise (S/N) of the images is very low, \textsc{statmorph} produces an error in the calculation, hence these cases can be easily discarded. 
In general, the effect of cosmological dimming, S/N and image resolution on morphological parameters has been extensively investigated at different redshifts and for a variety of surveys \citep[e.g.,][]{lotz11,petty14,whitney21,westcott25,sazonova25}. 
Due to the different effects of the complications listed above, it is worth discussing the effects on asymmetry separately from the effects on Gini and M$_{20}$.

According to \citet{lotz04}, the asymmetry A, which requires image subtractions and corrections for the background noise signal, becomes less robust at low S/N$_{\rm pixel}$. This parameter is also affected by the degradation of the image resolution, even though this effect is not as significant as the dependence on signal-to-noise \citep{thorp21}. 
In particular, the asymmetry tends to be underestimated both at low S/N$_{\rm pixel}$ and low effective resolution (i.e., the number of resolution elements sampling the galaxy, \Citealt{bottrell19}, \Citealt{thorp21}, \Citealt{sazonova24}). 
However, \citet{lotz11} found that by setting a lower limit on the S/N$_{\rm pixel} \geq 2$ in HST F160W images of galaxies at $z \leq 2$, the asymmetry is robust and not affected by systematic biases. Similarly, \citet{conselice_arnold09} find that asymmetry measurements are accurate also at higher redshifts (between $2$ and $6$), hence they do not apply any corrections to their measured values.
More recent simulations by \citet{sazonova25} have shown that at S/N$_{\rm pixel} \geq 3$ and with the F444W resolution, the underestimation is relatively small with a maximum bias of $0.05$. 
Finally, \citet{westcott25} have performed a series of simulations to test the effect of surface brightness dimming on the measurement of morphological parameters of galaxies at $z>5$ using JWST mock images. 
They estimated the morphological parameters of the galaxies at their photometric redshifts (with a similar procedure to the one adopted in this paper) and compared the results to those obtained when artificially redshifting the galaxies to a higher redshift. Their main finding is that the asymmetry does not differ significantly when the galaxies are simulated at higher redshifts, provided that the signal-to-noise is sufficiently high. In particular, at redshift $z>8$, it might be slightly underestimated by $2.5 \%$ on average.

On the other hand, Gini and M$_{20}$ are more robust against decreasing S/N compared to the asymmetry A, changing less than $10\%$ ($\delta_{Gini} < 0.05$ and $\delta_{M_{20}} < 0.2$) for an average S/N$_{\rm pixel}$ $>2$ if the spatial resolution is better than $1$ kpc \citep{lotz04}. Their reliability down to average S/N$_{\rm pixel}$ $\geq 2$ has been independently confirmed by other works \citep[i.e.,][]{holwerda11,lisker08,sazonova25,petty14,yao23,salvador24}.  
These parameters are instead more sensitive to decreasing spatial resolution. In particular, the former is slightly underestimated at low resolution, while the latter can be overestimated. However, 
both are stable when the segmentation map of the detected source is covered by a sufficient number of pixels (approximately $> 40$), which is always satisfied by the mergers in our sample, as we mostly trace pre-coalescence systems with a more extended structure than isolated galaxies or post-mergers. 
Secondly, we use a combination of Gini and M$_{20}$. Since Gini$+0.14\times$ M$_{20}$ measures the perpendicular distance in the G-M$_{20}$ plane, the resolution dependencies of the two parameters mostly cancel out, resulting in a stable measurement for all the galaxies with good S/N$_{\rm pixel}$ \citep{lotz04}. 

Overall, previous studies consistently showed that above a minimum spatial resolution, the S/N$_{\rm pixel}$ can be used to discriminate between reliable and unrealiable morphological parameters estimations, hence to select galaxy samples with more robust structural properties. There might still be a slight underestimation of the asymmetry parameter, but we prefer not to apply any corrections to the measured values as they would be very uncertain and highly inaccurate (being derived from simulations). 
With this choice, it will also be easier to compare our analysis to other works (both at lower and similar redshifts) using the standard asymmetry definition and merger threshold \citep[e.g.][]{conselice09,dalmasso24,westcott25}.

\subsection{Final sample selection}\label{sec:final_sample_selection}

We follow a similar approach to the studies presented above, and require a minimum S/N$_{\rm pixel}$ of $3$ for our sample (as derived from \textit{statmorph}), slightly more conservative than in \citet{lotz04} and \citet{lotz11}. 
We have verified that this choice works better at isolating problematic fits than applying a simple cut based on the S/N of the total photometry of the galaxy as done in \citet{dalmasso24}. In particular, this condition removes all the measurements where \texttt{statmorph} yields unphysical, negative asymmetries. 
With this condition, we obtain a final sample of $1233$ galaxies, $\sim56 \%$ of the initial sample. Their redshift distribution is shown in Fig. \ref{fig:histograms_selected_sample} (red line). Despite this further selection, the shape of the distribution, the redshift limits ($5<z<14$), and the median redshift of the final sample (z$_{\rm median}$ $=6.0$), are similar to those of the original spectroscopic sample. 

In Fig. \ref{fig:histograms_selected_sample}-\textit{bottom} we show the distribution of M$_{ \rm UV}$ of our final sample as a function of redshift. 
Similarly as above, the distribution of M$_{\rm UV}$ is not significantly different between the original sample and the final selected one, implying that the S/N$_{\rm pixel}$ cut in F444W does not introduce systematic biases on M$_{\rm UV}$. The median M$_{\rm UV}$ of the final sample is $-19.3$, and it is not a strong function of redshift. 
Our final sample is on average $1$ magnitude fainter than the sample of \citet{dalmasso24}, due to their very strict requirements on the total photometric S/N of the sources (S/N$_{\rm total}$ $>7$) and on their apparent magnitude ($m_{AB} < 28.87$), hence we are more complete toward fainter objects. Notably, our M$_{\rm UV}$ distribution is similar to the entire DJA spectroscopic catalog from \citet{heintz25}, which has a median M$_{\rm UV} \simeq -19.6$ mag (standard deviation of $1$ mag). Our sample is instead shifted toward brighter magnitudes on average compared to purely photometric JWST samples in the literature. For example, considering the photometric catalog of \citet{endsley24} in JADES, one of the deepest JWST imaging surveys, our galaxies are brighter in M$_{ \rm UV}$ by approximately $0.6$ magnitudes at $z=6$. 
We have also checked that M$_{ \rm UV}$ does not show a correlation with S/N$_{\rm pixel}$. In particular, galaxies with S/N$_{\rm pixel}$ lower and higher than $3$ do not have significantly different M$_{ \rm UV}$ distributions, as the M$_{ \rm UV}$ does not take into account the redshift of the source and its physical size. For this reason, M$_{ \rm UV}$ is not as effective as the S/N$_{\rm pixel}$ to isolate problematic fits of the morphological parameters.

\subsection{Derivation of stellar masses and SFRs from SED fitting}\label{sec:SED_fitting} 

\begin{figure}[t!]
    \centering
    \includegraphics[angle=0,width=0.95\linewidth,trim={0.cm 0.cm 0.cm 0.cm},clip]{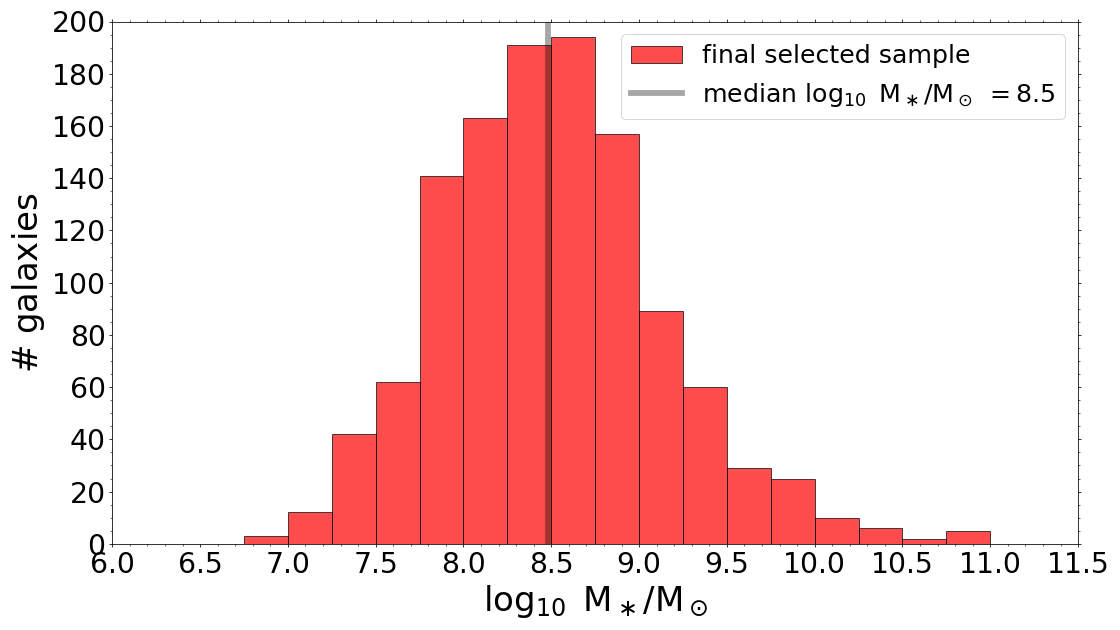}
    \vspace{-0.3cm}
    \caption{Histogram of the stellar mass distribution of the final selected, spectroscopic sample. The median stellar mass of the sample is highlighted with a vertical black shaded line. 
    }\label{fig:mass_histogram}
    \vspace{-0.3cm}
\end{figure}

We derived two fundamental galaxy properties, namely the stellar mass and the SFR, from an SED fitting procedure using the available photometric data covering the rest-frame UV and optical emission of the galaxies, following the methodology of \citet{llerena26}.  
In brief, we adopted the ASTRODEEP-JWST photometric catalog of \citet{merlin24}, which includes NIRCam + HST multiband total photometry covering all the five extragalactic deep fields considered in this work, and spanning the wavelength range between $0.4$ and $4.4$ $\mu m$. This range includes the following $16$ filters in order of wavelength: F435W, F606W, F775W, F814W, F090W, F105W, F115W, F125W, F140W, F150W, F160W, F200W, F277W, F356W, F410M, and F444W. The object detection is made with SExtractor v2.8.6 on a weighted F444W + F356W image, and total photometry is derived in each photometric band with the A-PHOT software \citep{merlin19}, with a measurement setup that is optimized for the faint and extended objects found in the high redshift Universe.

We fitted the photometry with BAGPIPES (version 1.2.0, \Citealt{carnall18}), adopting \citet{gutkin16} stellar population templates, and fixing the redshift to the spectroscopic value. We adopted a non-parametric SFH as in \citet{leja19}, according to which a constant SFR is fit in seven bins of increasing lookback times. In the first four bins, the SFR is calculated within the following ranges: 0–3, 3–10, 10–30, and 30–100 Myr, while the last four bins are logarithmically spaced between $100$ Myr and the maximum allowed age (corresponding to $z=20$). The stellar metallicity was free to vary in the range $0.01$-$0.5$ Z$_\odot$. We adopted a \citet{calzetti00} attenuation law with A$_{\rm V}$ ranging from $0$ to $2$ magnitudes. Finally, a nebular component was included based on CLOUDY photoionization models \citep{ferland17}, and the ionization parameter log(U) was free to vary between $-4$ and $0$. The IGM attenuation is modeled according to \citet{inoue14}.  

For the scope of our current project, deriving the SFRs from the photometry presents two main advantages compared to line based estimates. First, most of the spectra are taken with the NIRSpec-MSA configuration, whose slitlets ($0.2\arcsec \times 0.46 \arcsec$) might trace only a small portion of the galaxies and are thus not representative of the whole systems, especially in the presence of merger features and extended interacting components. Even calibrating the spectra with the available photometry, the line based properties rely on the assumption that the gaseous emission has the same spatial distribution as the stellar continuum emission, which might not hold for all the galaxies. Using the total photometry for the estimation of physical properties allows us to minimize these biases. 
Secondly, our SED fitting procedure allows us to compute SFRs averaged over timescales of $3$ Myr, $10$ Myr, $30$ Myr, and $100$ Myr. As shown by \citet{llerena24}, the SFRs averaged over the last $10$ Myrs agree well with those derived from optical recombination lines. Using the different timescale estimates, we can define a burstiness parameter as the ratio of a short timescale SFR over the $100$ Myr averaged SFR, as SFR$_{\rm t_{short}}$/SFR$_{100 \rm Myr}$, where t$_{short}$ varies between $3$ Myr and $10$ Myr. We will use this parameter in Section \ref{sec:burstiness}. 

The stellar masses of our sample range between $\log_{10}$ M$_\ast$/M$_\odot$ $=7$ and $11$, with a median value of $8.5$ ($1\sigma =0.6$), as shown in Fig. \ref{fig:mass_histogram}. 
This is $\sim 0.6$ dex higher than the median stellar mass that can be reached by JWST with a purely photometric selection, such as the one performed by \citet{duan25}, suggesting that our spectroscopic selection and the S/N$_{\rm pixel}$ threshold trace on average more massive galaxies compared to a photometrically selected sample at the same cosmic epoch. 
We will further analyse the SFRs in Section \ref{sec:SFR_enhancement}.

\subsection{Merger identification}\label{sec:merger_identification}

\begin{figure}[t!]
    \centering
    \includegraphics[angle=0,width=0.95\linewidth,trim={0.cm 0.cm 8.5cm 0.cm},clip]{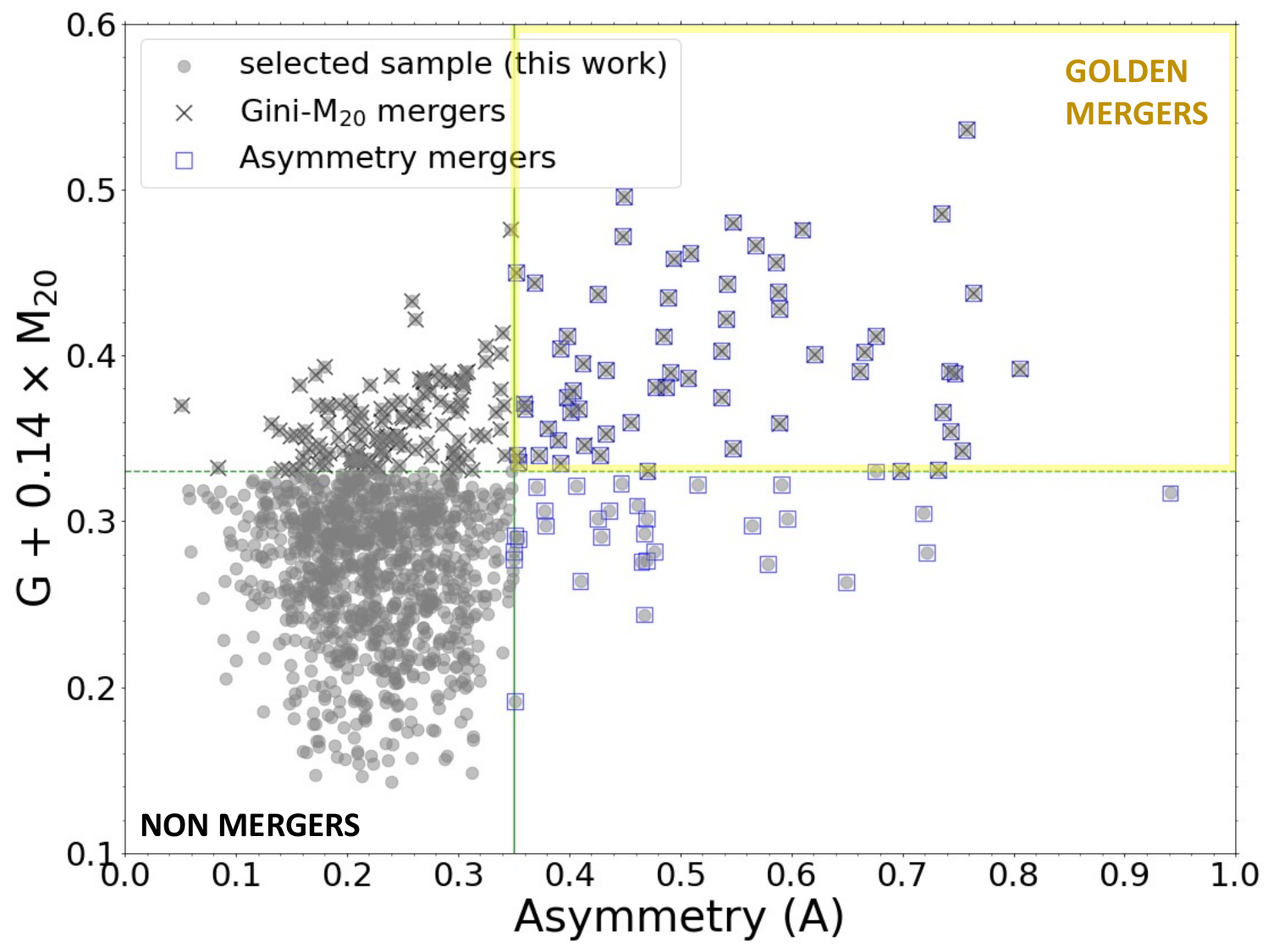}
    \vspace{-0.3cm}
    \caption{Diagram of the Asymmetry (A) vs f(G,M$_{20}$) for our selected sample, where f(G,M$_{20}$) $=$ Gini $+ 0.14 \times $ M$_{20}$. The two merger criteria $A > 0.35$ and f(G,M$_{20}$) $>0.33$ (as defined in Equation \ref{eq:merger_criterion}) are shown with a vertical continuous and horizontal dashed green line, respectively. Non merger galaxies are highlighted with simple gray circles. On top of them, we show a black cross for Gini-M$_{20}$ mergers and a blue empty square for asymmetry mergers. The golden merger region is delimited by a yellow rectangular line. 
    }\label{fig:final_sample_selection}
    \vspace{-0.3cm}
\end{figure}

We use the morphological parameters introduced in Section \ref{sec:morphological_parameters_derivation} to classify galaxies as mergers and non-mergers. We follow a similar approach to our previous work in \citet{dalmasso24}. This is a combination of the merger classification criteria adopted in \citet{lotz04} and \citet{conselice03}, which are based, respectively, on the Gini and M$_{20}$ parameters, and on the asymmetry. %  (A > 0.35). 
We thus classify a galaxy as a merger when both the two following conditions are satisfied : %Our primary merger selection is thus based on the following conditions :
\begin{equation}\label{eq:merger_criterion}
\begin{split}
A > 0.35  \\
G + 0.14 \times M_{20} > 0.33 
\end{split}
\end{equation}

This corresponds to the gold sample criterion defined in \citet{dalmasso24}, and represents our primary merger selection. Combining the two conditions maximizes the robustness of the merger selection. These limits have been validated at $z>5$ in \citet{dalmasso24}, and are based on the evidence that the intrinsic morphological parameters of mergers and normal isolated galaxies do not significantly evolve with redshift on average \citep{treu23,vulcani23}, allowing us to use the same thresholds that are adopted in the original studies at $z \leq 3$. We can visualize the distribution of the Asymmetry, Gini, and M$_{20}$ parameters of our sample in Fig. \ref{fig:final_sample_selection}, where the two conditions in Equation \ref{eq:merger_criterion} are highlighted with green lines and divide the scatter plot in four parts. We find that $80$ galaxies are classified as mergers according to both conditions, occupying the top right quarter in Fig. \ref{fig:final_sample_selection}. 

We additionally consider two merger subsets, dubbed the asymmetry and the Gini-M$_{20}$ mergers, which are identified by applying respectively the first and second condition of Equation \ref{eq:merger_criterion}. We create these additional samples as we think that they likely probe different merger types and different merger observability timescales, as we will explain in the following section.  We find in total $92$ galaxies satisfying the $A > 0.35$ condition, while $186$ galaxies fall in the category of Gini-M$_{20}$ mergers. $1016$ galaxies ($\sim80 \%$ of the sample) are instead non-mergers. These classes of galaxies are highlighted with different symbols in Fig. \ref{fig:final_sample_selection}. Example F444W cutouts of galaxies for each morphological class are shown in Fig. \ref{fig:example_cutouts} in appendix \ref{appendix:cutouts}.

We note that the merger criteria defined in Equation \ref{eq:merger_criterion} based on morphological parameters select galaxies with major morphological disturbances and with bright off-centered components, which are the best candidates of ongoing mergers following the indications of previous simulations \citep{conselice07,lotz11,snyder15,snyder19}. In this work, for our goals, we identify morphological disturbed galaxies as mergers, and use `merger candidate fraction' as a synonym of `true merger fraction'.

This merger definition has some limitations. First, f(G,M$_{20}$) and the asymmetry might not be enhanced (compared to normal isolated galaxies) during all phases of the interaction. Secondly, the amount of their enhancement depends on a variety of factors, including impact parameters, mass ratios, gas fractions, and spin of the colliding galaxies \citep[e.g.,][]{dimatteo08}. 
Moreover, they might not identify the majority of late stage interactions and post mergers because interacting features tend to rapidly fade after the coalescence (\Citealt{lotz08_simul}, see also discussion in \Citealt{pawlik16}). 
Finally, we remark that our approach differs from the standard techniques based only on the analysis of morphological parameters, as we also perform a visual inspection (as discussed in Section \ref{sec:morphological_parameters_derivation}) of the galaxies aimed at maximizing the purity of the merger sample.

\subsection{Merger timescales}\label{sec:merger_timescales}

\begin{figure}[th]
    \centering
    \vspace{-0.44cm}
    \includegraphics[angle=0,width=1\linewidth,trim={0.cm 0.cm 0.cm 0.cm},clip]{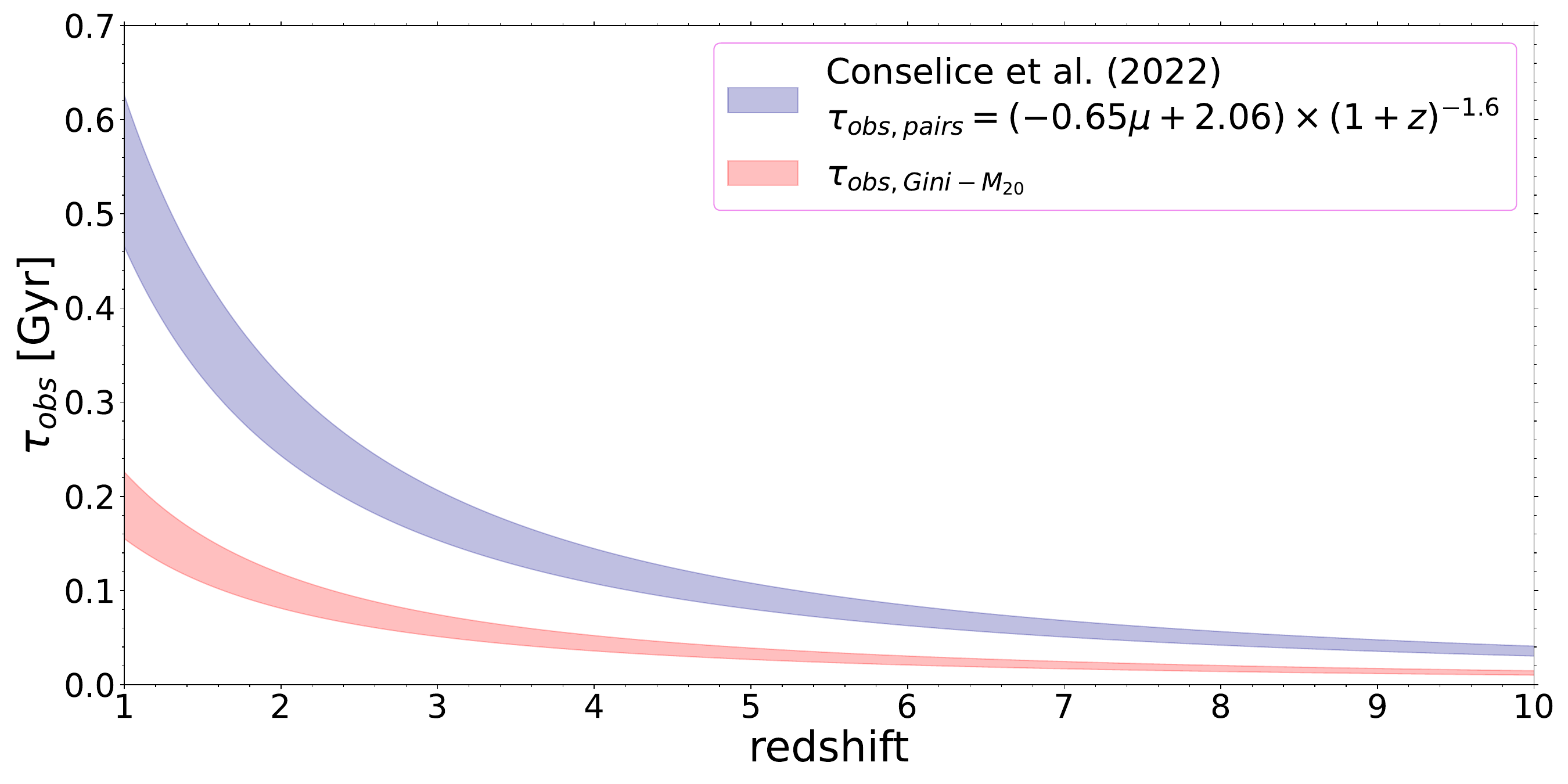}
    \vspace{-0.74cm}
    \caption{Merger observability timescales $\tau_{obs}$ in the redshift range $0$-$10$ that we assume in this work for the close pair method (from \Citealt{conselice22}, shaded dark blue region) and for the Gini-M$_{20}$ merger criterion (salmon region). 
    The blue shaded region encompasses the $\tau_{obs}$ expected for mass ratios $\mu$ between $0.25$ and $1$, while the salmon shaded region encompasses mass ratios $\mu$ ranging between $0.01$ and $1$. 
    }\label{fig:merger_timescales}
    \vspace{-0.1cm}
\end{figure}

In order to convert the observed merger fraction to a physical rate at which galaxies merge, it is essential to understand how the merger dynamical time evolves with cosmic time and what is the observability timescale probed by each specific tracer. 
The merger dynamical time $\tau_m$ is the timescale by which two merging galaxies finally coalesce into a single nucleus. This timescale decreases with redshift, such that galaxies in the early Universe merge faster than those at lower redshifts \citep{snyder17}. 
An analytic function describing this redshift variation has been recently derived by \citet{conselice22}, who used IllustrisTNG 300-1 simulations to study at different redshifts how long it takes at different redshifts for a close pair to finally merge. (see also \Citealt{ferreira20}). 
They found that $\tau_m$ also depends on the mass ratio $\mu$ of the merging galaxies, such that major mergers (higher $\mu$) are faster than minor mergers (lower $\mu$). 

In particular, they derived the following expression for $\tau_m$, which assumes a projected pair separation of $\sim 30$ kpc : 
\begin{equation}\label{Eq:merger_timescale_conselice22}
\tau_m(z)=a \times (1+z)^b ,
\end{equation}
where $a=-0.65 \pm 0.08 \times \mu + 2.06 \pm 0.01$, and $b=-1.60 \pm 0.01$. 
Slightly different parameters, with an exponent $b=-2.0$ and $a=2.4$, have been proposed by \citet{snyder17}, but they yield results that are not significantly different than our assumed values from \citet{conselice22}. The modeled evolution of the dynamical merger timescale is also similar to the typical halo dynamical timescale, $0.1t_\mathrm{H}(z)$, where $t_\mathrm{H}(z)$ is the Hubble-time (i.e. age of the Universe) at redshift $z$. However, using the typical halo dynamical timescale would yield slightly longer $\tau_m$ at $z<1$ and shorter $\tau_m$ at $z>1$ compared to the function in equation \ref{Eq:merger_timescale_conselice22}.  

We note that $\tau_m$ depends on the maximum physical distance considered for the photometric pairs in the simulation. Selecting interacting pairs closer (wider) than $30$kpc projected, we should consider shorter (longer) merger timescales. 
According to \citet{lotz11}, assuming baryonic mass ratios $\mu$ in the range $0.25$-$1$ (i.e., major mergers), the close pair method at low redshift ($0<z<1$) traces $\tau_m$ of the order $\sim 330$ Myr when adopting small projected separations of $5$ kpc $h^{-1} < R_{proj} < 20$ kpc $h^{-1}$, $600$ Myr for pairs with $10$ kpc $h^{-1} < R_{proj} < 30$ kpc $h^{-1}$, and $\geq 1.5$ Gyr for larger separations of $\gtrsim 100$ kpc $h^{-1}$. 

If we consider mergers identified through their morphology, several studies have shown that different merger conditions and morphological parameters are sensitive to different phases of the interaction, typically shorter than the merger dynamical time defined above \citep[e.g.,][]{lotz08_simul,lotz10,conselice09,conselice22}. This phenomenon can be parametrized by the merger observability timescale $\tau_{obs}$, defined as the timescale during which a merger can be identified by a particular observational method. However, $\tau_{obs}$ of the method only plays a role if it is shorter than the intrinsic merger event timescale. For the close pair method, we can assume that $\tau_{obs,pairs}$ is equal to $\tau_m$ defined above and analyzed in simulations. Its redshift variation can be seen in Fig. \ref{fig:merger_timescales} (dark blue curve).

Regarding morphological merger criteria, the Gini-M$_{20}$ condition is sensitive to minor and major mergers according to \citet{lotz11}, probing mass ratios of the colliding galaxies between $1:1$ and $1:10$ (mass ratio $\mu$ between $1$ and $0.1$), over timescales of $\sim 200$ Myr (at $z \sim 1$), independent of the mass ratio. This is indeed the phase when the two galactic nuclei are enclosed in a common envelope, thus they are more likely to be identified within the same segmentation map, increasing both Gini (because of unequal flux distribution) and M$_{20}$ (because of bright off-centered structures).

The asymmetry condition $A > 0.35$ is sensitive to major mergers, in the mass ratio range between $1:1$ and $1:4$ (mass ratio $\mu$ between $1$ and $0.25$), as mergers with lower mass ratios produce fainter and undetectable asymmetric features \citep{conselice03,conselice07,lotz11}. In this case the observability timescale is more difficult to model as it is a strong function of the gas fraction and of the mass ratio of the colliding galaxies \citep{lotz11}. The strong dependence on $f_{\rm gas}$ arises because more gas-rich mergers are more likely to form strong tidal features and large scale disturbances, which also last longer than in gas poor mergers \citep{lotz10}.  
Previous studies have attempted to estimate $\tau_{obs}$ related to an enhanced asymmetry parameter ($\tau_{obs,Asymmetry}$).
According to \citet{lotz11}, for $f_{\rm gas} = 0.4$ and $\mu \simeq 1$:$3$, the merger observability timescale is similar to the one traced by the photometric pair method for separations $\leq 30$kpc (i.e., $\sim 600$Myr at $z\sim1$). 
In the simulations presented by \citet{conselice07}, the $A > 0.35$ condition is not satisfied over the full time of the merging process, but only for $\sim 2/3$ of the merger duration, for a total observability time of $\tau_{obs,A>0.35}$ $=0.4$ Gyr on average at $z \simeq 1$. In both cases, this timescale is significantly longer compared to $0.2$ Gyr traced by the Gini-M$_{20}$ condition. Because of this considerations, $\tau_{obs,A>0.35}$ at $z \sim 1$ is likely comprised between $\tau_{obs, Gini-M_{20}}$ and $\tau_{obs, pairs}$. 
If we combine the two criteria in Equation \ref{eq:merger_criterion}, we are thus tracing major mergers with observability timescales of $\sim 200$ Myr at $z \sim 1$, because Gini $+ 0.14 \times $ M$_{20} > 0.33$ sets the strictest upper limit on the merger observability timescale for all mass ratios, and, at the same time, $A > 0.35$ selects only major mergers ($\mu > 0.25$).

At higher redshifts, there are no specific studies focused on the evolution of the observability timescales. Therefore, in this work, we estimate them at $z>1$ using the following reasoning. 
For $\tau_{obs, Gini-M_{20}}$, we consider the merger timescale $\tau_m$(z) as described by equation \ref{Eq:merger_timescale_conselice22}, and we rescale it by a constant factor as $\tau_{obs, Gini-M_{20}}$(z,$\mu$) $=$ $\tau_m$(z,$\mu$) $\times$ $C$. The constant factor $C$ is anchored to $z \sim 1$ and is set to $1/3$, ensuring that $\tau_{obs, Gini-M_{20}}$(z$=1$,$\mu$) $\simeq$ $200$ Myr, in agreement with Lotz et al. (2011). In practice, we assume that, at all redshifts, the Gini-M$_{20}$ criterion traces merger timescales that are one third of that probed by the close pair method with $30$ kpc separation. The final $\tau_{obs, Gini-M_{20}}$(z) can be seen in Fig. \ref{fig:merger_timescales} for a range of $\mu$ from $0.1$ to $1$ (characteristic of minor + major mergers), and it follows a similar decreasing trend as $\tau_m$(z), with smaller timescales probed as we go to higher redshifts, ranging from $\sim 30$ Myr at $z=5$ to $\sim 10$ Myr at $z=10$. 
A similar evolution of the merger timescale can be applied to the gold merger criterion, which thus probes major mergers with $\tau_{\rm obs}$ $< 30$ Myr at $z>5$. 
We emphasize that the estimation of this parameter is very uncertain. However, it is physically reasonable to assume that it cannot be larger than the merger timescale itself and than the characteristic $\tau_{obs}$ of close photometric pairs, which would set an upper limit of $\sim 100$ Myr at $z>5$. 

Considering the asymmetry observability timescale, we should expect a milder decrease with redshift at $z>1$ compared to $\tau_{obs, Gini-M_{20}}$ because the increasing gas fraction of typical star-forming galaxies \citep{daddi10} partially balances the shorter merger duration in the early Universe. Following the indications of previous studies at lower redshifts, it is reasonable to assume that $\tau_{obs, A > 0.35}$ at $z>5$ is likely comprised between $\tau_{obs, Gini-M_{20}}$ and $\tau_{obs, pairs}$. 
However, given its strong dependence on the mass ratio and gas fraction, whose evolution at the cosmic epochs explored in this study is also very uncertain, we do not attempt to derive any quantitative estimates of this timescale, and we do not use it for the calculation of merger rates in the following part of the paper.

\section{Results}\label{sec:results}

\subsection{Evolution of the merger fraction at z > 5}\label{sec:merger_fractions}

\begin{figure*}[t!]
    \centering
    \includegraphics[angle=0,width=\linewidth,trim={0.cm 1.8cm 2.5cm 0.cm},clip]{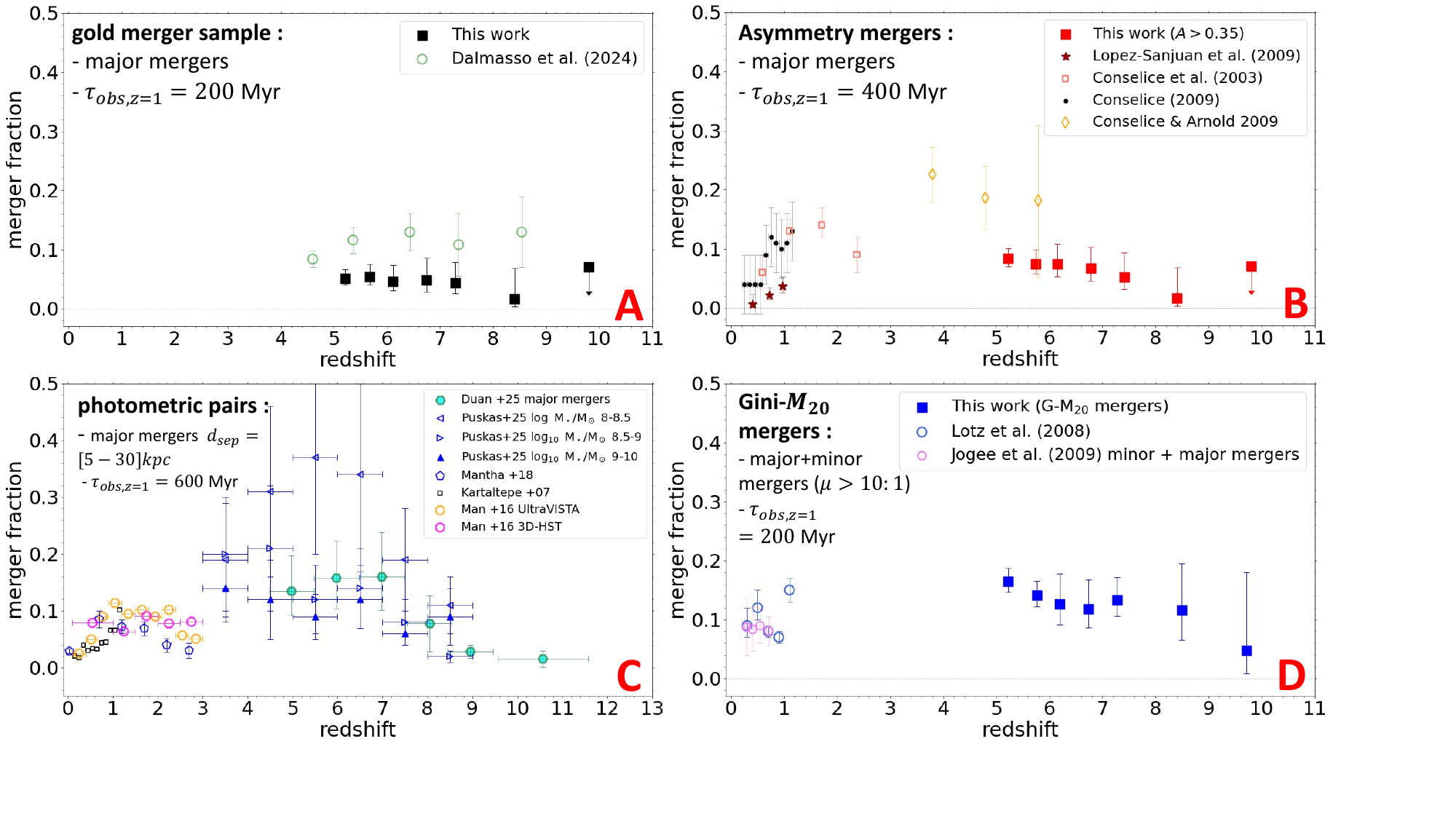}
    \vspace{-0.44cm}
    \caption{Merger fraction f$_m$ as a function of redshift for different merger identification methods : panel A) gold merger sample; panel B) asymmetry mergers satisfying $A > 0.35$; panel C) mergers identified through the photometric pairs; panel D) Gini-M$_{20}$ mergers. In each panel, our data is compared to several works in the literature at lower and similar redshifts adopting a similar merger identification technique. Golden mergers are compared to \citet{dalmasso24}, asymmetry mergers to \citet{conselice03,conselice09,conselice_arnold09,lopez-sanjuan09}, Gini-M$_{20}$ mergers to \citet{lotz08_obs,jogee09}. The results with the photometric pair method are from  \citet{duan25,puskas25_mergers,mantha18,kartaltepe07,man16}.  
    }\label{fig:merger_fraction}
    \vspace{-0.3cm}
\end{figure*}

In this section we calculate the fraction of mergers in our sample, and investigate how it varies across our redshift range and how it depends on other galaxy properties. 
We define the merger fraction in $7$ bins of increasing redshifts as : 
f$_m = \# mergers_{z-bin}$ / $\# galaxies_{z-bin}$, with $1\sigma$ uncertainties estimated using the % low number 
statistics of Gehrels et al. (1986). The redshift bins are defined as $5$-$5.5$, $5.5$-$6$, $6$-$6.5$, $6.5$-$7$, $7$-$8$, $8$-$9$, and $9$-$12$, where the bins at higher redshifts are larger in order to retain a good statistics and a minimum number of $15$ galaxies per bin.

We perform this analysis for different morphological indicators and merger criteria. 
The results are presented in the four panels of Fig. \ref{fig:merger_fraction}, which show the redshift evolution of the merger fraction using the golden merger sample (panel A), the asymmetry merger sample (panel B), and the Gini-M$_{20}$ merger sample (panel D). For comparison purposes, we also show the merger fraction obtained in the literature with the close photometric pair method (panel C). In panels A,B, and D, our results (highlighted with big filled squares) are plotted with those obtained with similar merger selections (or tracing mergers of the same type) in other works at similar and lower redshifts, down to $z=0$ \citep{dalmasso24,conselice03,conselice09,conselice_arnold09,lopez-sanjuan09,lotz08_obs,jogee09,duan25,puskas25_mergers,mantha18,kartaltepe07,man16}. The comparison with these works will be discussed more in detailed in the following section.

Focusing on our results, we find that the golden merger fraction is approximately constant at redshifts $5<z<8$ with a median f$_{\rm m,golden}$ $= 4.8 \%$. It slightly decreases at earlier cosmic epochs ($z>8$), even though in the last two bins, where we have poorer statistics, f$_{\rm m,golden}$ is still consistent with the median value (Fig. \ref{fig:merger_fraction}- panel A). 
For the asymmetry mergers, f$_{\rm m,asymmetry}$ varies between $\sim 8 \%$ and $5 \%$ from redshift $5$ to $8$, with a median f$_{\rm m,asymmetry}$ $= 7.4 \%$. It decreases at higher redshifts, even though the uncertainties are significantly higher (Fig. \ref{fig:merger_fraction}- panel B). 
Finally, for the Gini-M$_{20}$ mergers, f$_{\rm m,Gini-M_{20}}$ ranges between $\sim 11 \%$ and $\sim 17 \%$ at $5<z<9$, while at $z>9$ it drops to $\sim 5 \%$ (Fig. \ref{fig:merger_fraction}- panel D). 
 
Overall, merger indicators based on morphological parameters suggest that the merger fraction remains approximately constant between redshift $5$ and $8$ and then slightly decreases at earlier epochs.  We find slight differences in normalization among different methods. For major merger tracers (i.e., asymmetry mergers and golden mergers), longer merger timescale indicators yield on average higher merger fractions than short timescale indicators.   
Similarly, at fixed observability timescale, minor merger indicators yield higher merger fractions than major merger ones. 
The slight decrease in the last two bins is affected by low number statistics and is not highly significant. We would need a larger sample to confirm this trend at $z>8$. However, we note that a similar decreasing trend of the merger fraction is also found with the photometric pair method f$_{\rm m,pairs}$ (see Fig. \ref{fig:merger_fraction}- panel C), using studies based on larger (photometric) samples \citep{duan25,puskas25_mergers,mantha18,kartaltepe07,man16}, corroborating this observed trend. 
We summarize our findings in Table \ref{table:results_merger_fractions} in the appendix \ref{appendix:tables}, where we specify the median stellar mass and the merger fraction in each redshift bin.

\subsection{Comparison to other studies}\label{sec:comparison_other_studies}
 
It is interesting to compare more in detail our findings to previous studies at lower and similar redshifts. We specify the redshift range, the stellar mass range, the imaging used, and the merger selection method of other works in Table \ref{table:comparison_works}. We notice that lower redshift studies before the advent of JWST are mostly based on HST images for the merger identification, or ground based SDSS data up to $z=0.2$. The merger selection is typically done in the rest-frame optical of the galaxies. It is also worth noticing that lower redshift studies probe in general a higher stellar mass range, typically higher than $10^{10}$ M$_\odot$, compared to the one explored at $z>5$ with JWST, with the exception of \citet{conselice03}, who probe masses down to $10^8$ M$_\odot$ at $z<3$. With this in mind, we now analyze more in detail the comparison with other studies for each specific merger tracer.

The most direct comparison for the asymmetry sample is with the studies by \citet{lopez-sanjuan09}, \citet{conselice03}, \citet{conselice09}, and \citet{conselice_arnold09}, who adopt a morphological criterion which is identical to our one based on the asymmetry parameter (i.e., A$>0.35$) to investigate mergers at $z \leq 6$.  These studies indicate that there is a sudden increase of the merger fraction from the local Universe to $z\simeq 2$, going from $\sim 3\%$ to $\sim 15 \%$. The merger fraction decreases again to $\sim 10\%$ at earlier epochs ($z=2$-$3$). This merger fraction is similar to what we find at redshifts $5<z<7$, before a further decrease of f$_{\rm m,asymmetry}$ below $10\%$ is observed at $z>7$ (Fig. \ref{fig:merger_fraction}-panel B).  
The higher merger fraction found in \citet{conselice_arnold09} at redshifts $3<z<6$ might be due to a combination of cosmic variance, poorer statistics, high stellar mass probed, and photometric band used. This is the highest redshift study (pre-JWST) on morphologically selected mergers, pushing this investigation up to $z \simeq 6$ with HST images (in z-band). As a result, their analysis is made within a smaller field of $11 arcmin ^2$ (the Hubble Ultra Deep Field) where deeper observations were available. 
Their sample comprises $133$ galaxies divided in $3$ redshift bins, while similar works at lower redshifts are based on larger samples of $\sim 800$, $\sim 1200$, and $\sim 22000$ galaxies, respectively in \citet{lopez-sanjuan09}, \citet{conselice03}, and \citet{conselice09}. Their sample is also biased toward more massive galaxies (M$_\ast \gtrsim 10^{10}$ M$_\odot$) compared to their previous analysis in \citet{conselice03}. Finally, in contrast to other works at lower redshifts, the z-band used for their merger identification probes the rest-frame UV emission of galaxies at $z>3$, which is more affected than the optical by inhomogeneous dust distribution and more recent star-formation.

For the Gini-M$_{20}$ mergers, we compare our results to similar studies performed at lower redshifts with HST images by \citet{lotz08_obs} and \citet{jogee09}. While the former study adopts the same condition of our Gini-M$_{20}$ merger sample, the latter uses a combination of visual and quantitative merger classification which traces minor+major mergers (down to $\mu$ of $0.1$) according to the authors.
Also in this case, we do not find a significant evolution of the Gini-M$_{20}$ merger fraction from $z \sim 0.5$ to $z \sim 9$, with f$_{\rm m,Gini-M_{20}}$ ranging between $5 \%$ and $15 \%$ over that cosmic epoch (Fig. \ref{fig:merger_fraction}-panel D). 

There are no comparable studies at lower redshifts that consider the combination of the asymmetry and Gini-M$_{20}$ merger criteria. Therefore, we can only compare our golden merger sample to the results of \citet{dalmasso24}. 
We find merger fractions that are systematically lower than in that work at the same redshift (Fig. \ref{fig:merger_fraction}- panel A). Presumably, this is due to the fact that no visual check of the merger morphology with removal of low redshift companions were done in the previous analysis, leading to an overestimation of the true merger fraction. Indeed, we have verified that skipping those intermediate visual checks in the merger selection procedure yields merger fractions that are consistent with \citet{dalmasso24}.

For completeness, we also analyzee the evolution of the merger fraction from the close pair method. 
In Fig. \ref{fig:merger_fraction}-panel C we plot the pair fractions (f$_p$) obtained in several studies assuming a maximum projected distance separation of $30$ kpc between the two galaxies. 
Previous HST based works in the CANDELS fields have constrained the pair fraction up to redshift $\sim 3$, finding an increase of f$_{\rm m,pairs}$ up to $z \simeq 1$, and a flattening from $z=1$ to $z=3$ around a value of $10 \%$. 
Recently, JWST observations suggest that there is a slight increase of f$_p$, with peaks of $15 \%$ to $30 \%$ around $z \sim 6$, which decreases again at $z>7$. 

Overall, different merger indicators and selection techniques indicate that the fraction of mergers is not strongly evolving with redshift from $z \simeq 1$ to $z\simeq 8$.
We find instead different normalizations in the merger fraction vs redshift relations. This depends on the characteristic timescale of the merger diagnostic, with lower merger fractions found for shorter timescale indicators. In particular, the golden merger criterion yields merger fractions approximately a factor of $2$ lower than the photometric pair selection. More abundant mergers are also found for minor+major merger tracers, such as the Gini-M$_{20}$ merger criterion, compared to major merger tracers. 

As the stellar mass is one of the main differences between studies performed at low redshift and at $z>5$ with JWST, we investigate better the effect of the galaxy stellar mass on the estimated merger fraction. In previous works, this is a very debated question. \citet{puskas25_mergers} find a strong dependence of the pair fraction on stellar mass, with lower mass galaxies showing a higher f$_{\rm m,pairs}$ at fixed redshift, which explains the broad scatter of f$_{\rm m,pairs}$ found in that study at $z>5$ (Fig. \ref{fig:merger_fraction}). On the other hand, considering morphological studies, \citet{conselice03} did not find a significantly different trend of f$_{\rm m,asymmetry}$ as a function of stellar mass at $z<3$. 
Because of this uncertainty, we have also tested possible variations of the merger fraction as we choose different M$_\ast$ lower limits for our dataset, similar to the approach of \citet{conselice03} and \citet{conselice09}. In particular, we have considered lower limits of $10^8$ M$_\odot$ and $10^9$ M$_\odot$, as cutting at even higher masses significantly reduces the available statistics. 
As a result, we find a similar redshift trend for the two datasets 
, with an approximately constant f$_{\rm m,golden}$ across the different cosmic epochs, as observed for the global sample. Most importantly, we do not find any significant difference of the average merger fraction between the low mass and high mass galaxies in all the redshift bins, thus the stellar mass likely does not produce a strong selection bias in the results. 
However, it is currently unknown whether this result can be extrapolated to the highest mass range typical of low redshift studies (M$_\ast > 10^{10}$ M$_\odot$), as we do not have a good statistics in that regime. We therefore suggest taking this potential stellar-mass-related effect into account when comparing studies conducted at significantly different redshifts.

\subsection{Testing the SFR enhancement in merging galaxies}\label{sec:SFR_enhancement}

\begin{figure}[t!]
    \centering
    \includegraphics[angle=0,width=0.97\linewidth,trim={0cm 0cm 0cm 0cm},clip]{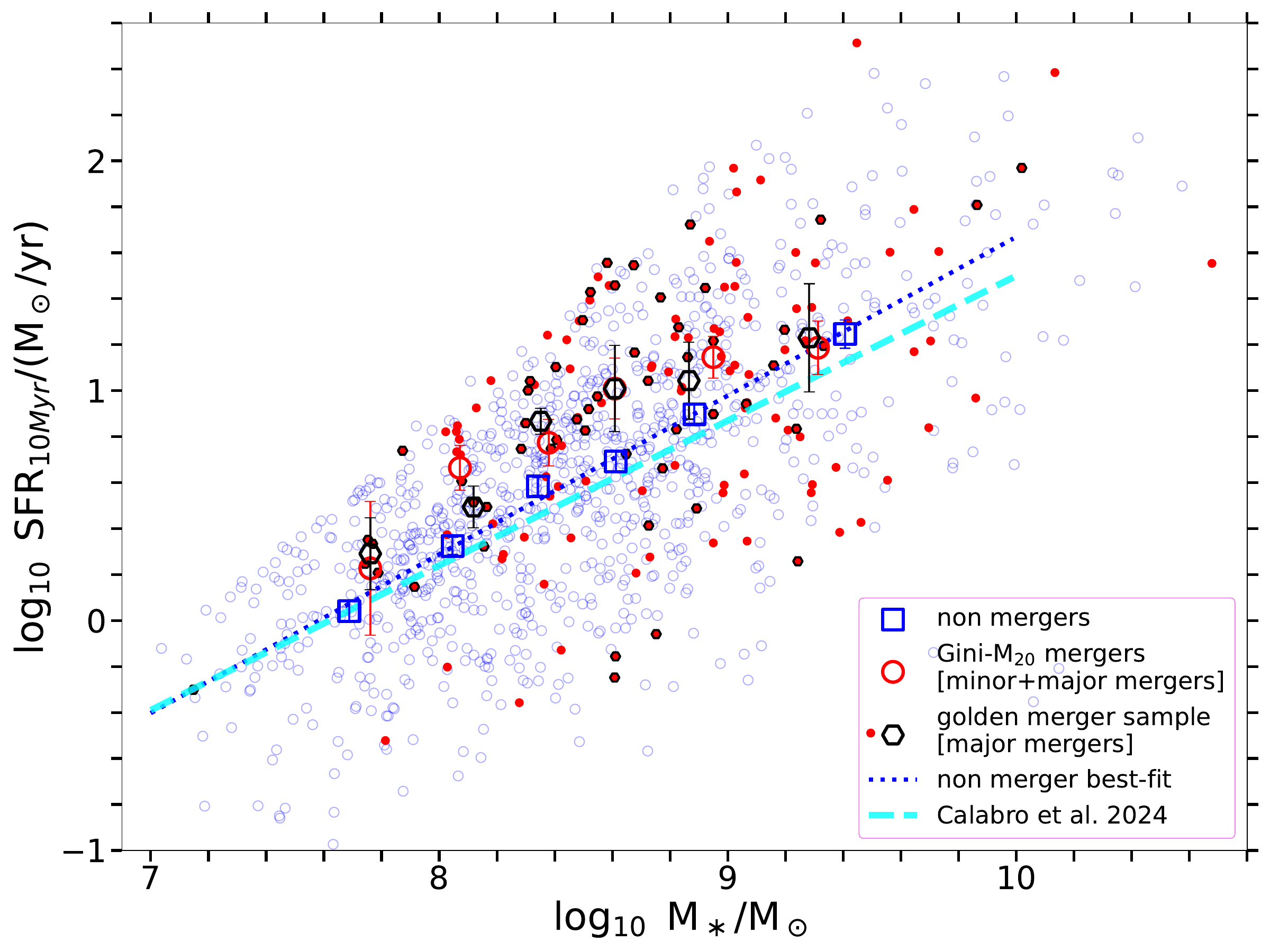}
    \vspace{-0.3cm}
    \caption{Stellar mass vs SFR diagram for mergers vs non merger galaxies, in the redshift range $5<z<8$. The golden merger sample and the Gini-M$_{20}$ mergers, tracing respectively the major and major+minor mergers, are shown in small black empty hexagons and red empty circles. The non-merger sample is shown with blue empty circles. Median SFRs (and uncertainties on the median) are calculated in six, equal populated redshift bins, and shown as bigger black hexagons, red circles, and blue squares, respectively for the golden mergers, Gini-M$_{20}$ mergers, and non-mergers. The best-fit linear relation ($y=m \times x + q$) to the non-merger bins is shown with a blue dotted line, with best-fit parameters $m=0.7 \pm 0.03$, and $q= -5.3 \pm 0.3$ . 
    The SFRs are averaged over $10$ Myr, thus comparable to those inferred from the H$\alpha$ line. The H$\alpha$-based Main Sequence of star-formation derived in \citet{calabro24} in the CEERS and GLASS surveys is highlighted with a cyan dashed line. 
    }\label{fig:SFR_mass}
    \vspace{-0.3cm}
\end{figure}

The main objective of the paper is to understand whether the merger process impacts the star-formation activity of early galaxies, as observed in the low redshift Universe. To this aim, we compare the stellar mass - SFR relation between our merger and non-merger galaxy subsets (Fig. \ref{fig:SFR_mass}). We primarily consider the gold merger sample (top-right quadrant in Fig. \ref{fig:final_sample_selection}), since they likely trace major mergers over a short timescale, where we expect to have the strongest effects on galaxy properties, if there are any. Moreover, this is the most robust merger sample, as resulting from the combination of three different morphological parameters. However, for completeness, we will also explore the star-formation enhancement in the Gini-M$_{20}$ merger sample, as they likely probe a larger range of mass ratios (down to $10$:$1$) with a similarly short observability timescale. 
We limit the analysis to the redshift range $5<z<8$, since the number of galaxies drops at $z>8$ resulting in poorer statistics (see Table \ref{table:results_merger_fractions}).
Moreover, we consider SFRs averaged over a timescale of $10$ Myr, as they are comparable to those derived in other observational studies from recombination lines (e.g., H$\alpha$). 

Dividing our samples in six equal populated stellar mass bins, we find a SFR vs M$_\star$ linear relation for non merging galaxies that is in agreement with the H$\alpha$-based star-forming Main Sequence (MS) derived by \citet{calabro24} from star-forming galaxies at similar redshifts in the CEERS and GLASS surveys (Fig. \ref{fig:SFR_mass}). This comparison also suggests that our initially selected sample does not have a significant bias compared to other JWST spectroscopic samples, in particular to the mass complete sample (down to an M$_\ast$ limit of $\log_{10}$ (M$_\ast$/M$_\odot$) $=7.8$ at $z \sim 6$) analysed in \citet{calabro24}. 

\begin{figure}[t!]
    \centering
    \includegraphics[angle=0,width=1\linewidth,trim={0.cm 3cm 19.5cm 0.cm},clip]{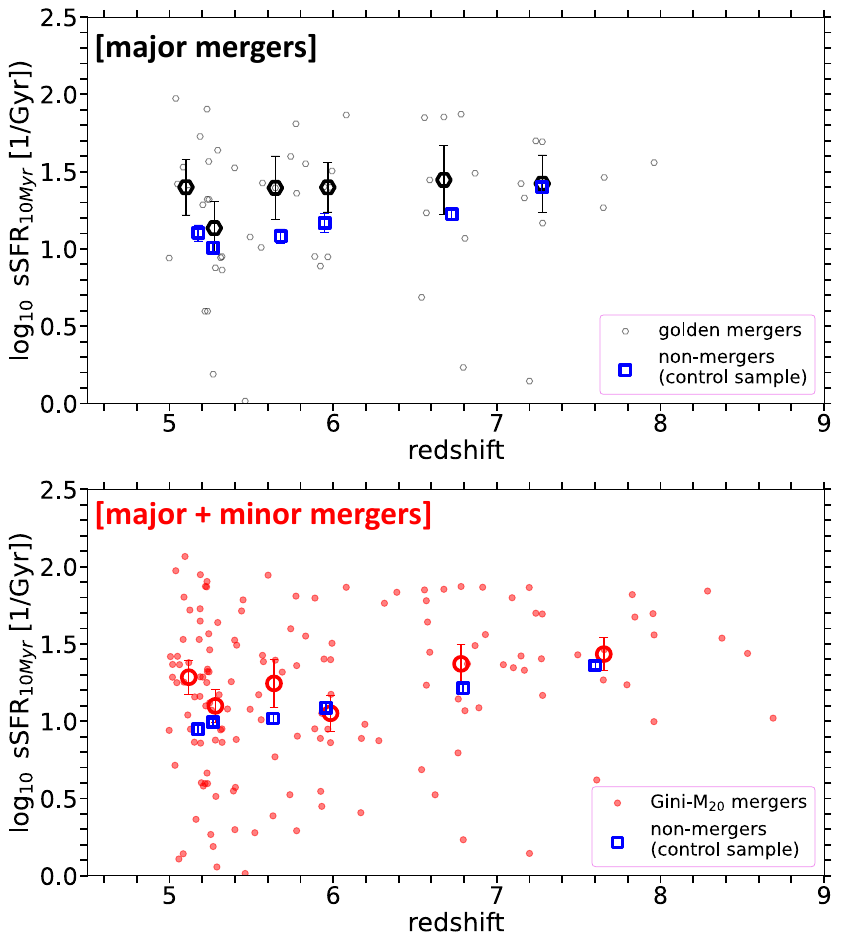}
    \vspace{-0.5cm}
    \caption{Specific SFR (sSFR) vs redshift diagram for the golden merger sample and Gini-M$_{20}$ mergers (respectively in the upper and lower panel), compared to a mass and redshift matched sample of isolated galaxies, using a SFR timescales of $10$ Myr. 
    }\label{fig:SSFR_redshift}
    \vspace{-0.35cm}
\end{figure}

While isolated galaxies agree well with the global MS relation, merging galaxies show systematically higher SFRs in all the stellar mass bins, except the most massive one above $\log_{10}$ (M$_\ast$/M$_\odot$) $=9.2$, where the three subsets have similar SFRs on average (Fig. \ref{fig:SFR_mass}). Considering the galaxies in the first $5$ bins in the range of M$_\star$ between $10^7$ and $10^9$ M$_\odot$, we find a SFR enhancement of $0.25 \pm 0.04$ dex for the golden and the Gini-M$_{20}$ merger sample, that is, a factor of $\sim 1.8$ higher compared to isolated galaxies. 
The different behaviour in the most massive bin could be due to timescale effects, or to different underlying physics. More massive mergers could be in a more advanced, post-coalescence merger stage, where the SFR has already reached its peak. Alternatively, more massive galaxies might be subject to stronger feedback preventing new star-formation.

To investigate the redshift dependence of this enhancement, we analyse the cosmic evolution of the sSFR. In particular, we calculate the median sSFR in $6$ equal-sized redshift bins for the merger subset and a control sample of isolated galaxies. 
This control sample was built by selecting $20$ times for each merger a corresponding non-merging galaxy (randomly) with a similar stellar mass (within $0.2$ dex) and a similar redshift (within $0.3$). 
We show this comparison in Fig. \ref{fig:SSFR_redshift} with a SFR averaged over the last $10$ Myr. 
We find that mergers have a significant enhancement of sSFR compared to isolated galaxies, with a median enhancement of $0.23 \pm 0.04$ dex for the golden subset, which does not significantly evolve with redshift. The Gini-M$_{20}$ mergers have a slightly lower but still significant sSFR enhancement of $0.13 \pm 0.04$ dex. The results shown in Fig. \ref{fig:SFR_mass} and Fig. \ref{fig:SSFR_redshift} for mergers and non-mergers are reported in Table \ref{table:results_SFR_enhancement}.

\begin{figure*}[t!]
    \centering
    \includegraphics[angle=0,width=0.86\linewidth,trim={0.cm 0cm 0cm 0.cm},clip]{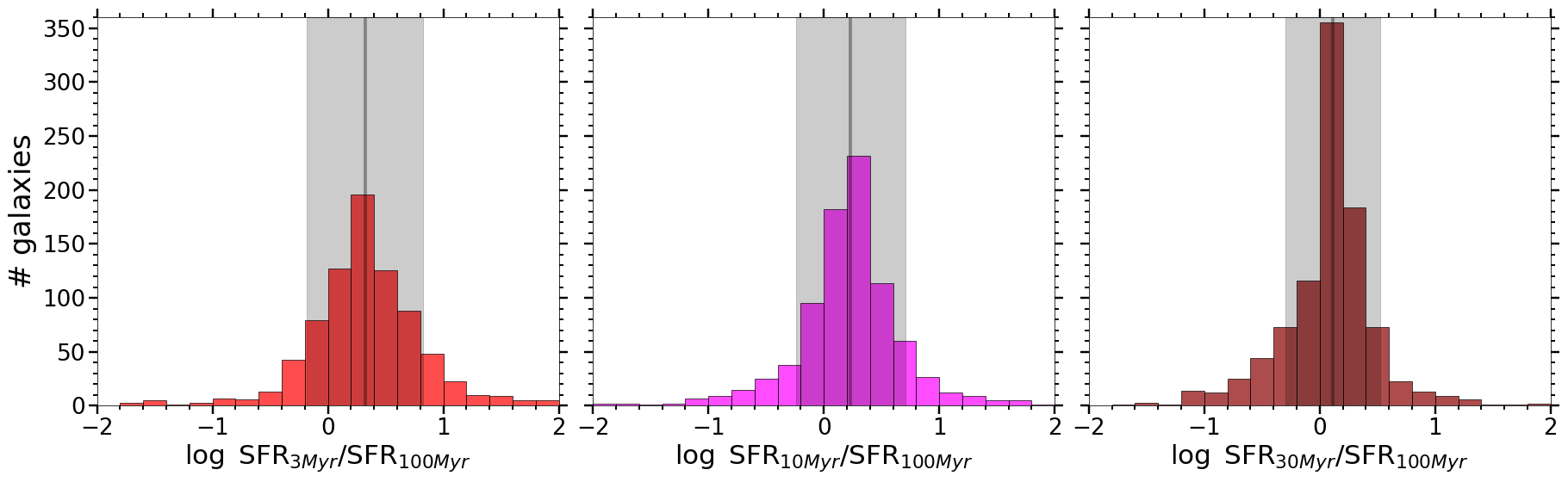}
    \vspace{-0.2cm}
    \caption{Burstiness histogram distribution of the entire sample selected in this work, for a timescale of $3$, $10$, and $30$ Myr (from left to right). The burstiness is shown in logarithmic scale, and we have limited the visualization to the range of $\log$ SFR$_{\rm timescale}$/SFR$_{100 Myr}$ between $-2$ and $2$. In each panel, the vertical black line indicates the median burstiness value, while the shaded gray area encompasses the $\pm 1 \sigma$ range of the distribution. 
    }\label{fig:burstiness_histograms}
\end{figure*}

Our findings are in line with \citet{puskas25_SFR}, who found a moderate sSFR enhancement in mergers (by $\sim10 \%$) in the JADES survey for galaxies with similar stellar masses and redshifts as this study, and using the smallest separation ($\leq 20$ kpc) to identify close photometric pairs. We find a slightly higher enhancement than in their study, which might be due to our selection method identifying more advanced merger phases where the two galaxies are more strongly interacting. 
We are also in agreement with \citet{duan25}, who find a SFR enhancement of a factor of $\sim 1.8$ for close pairs at the smallest projected separation ($\leq 20$ kpc) between redshift $4.5$ and $8.5$, while no significant enhancement is detected at larger separations. Finally, the non detection of a significant SFR enhancement in mergers in \citet{dalmasso24} might be due in part to their larger uncertainties (because of poorer statistics), and in part to the possible low-redshift contaminants discussed in Section \ref{sec:comparison_other_studies}, which might affect the estimation of physical parameters from SED fitting.
However, we note that our small effect would still be consistent with their $3\sigma$ limits. 

Finally, works at lower redshifts have investigated the merger induced SFR enhancement mainly in massive galaxies ($\log$ M$_\ast$/M$_\odot$ $> 9$-$10$). To ensure a fair comparison, we focus on studies probing advanced merger phases similar to those targeted by our quantitative selection.
In the local Universe ($0.01<z<0.2$), using the SDSS survey, \citet{ellison13} and \citet{patton13} find an average SFR enhancement of a factor of $\sim 3.5$ and $\sim 3$, respectively, in late-stage mergers (identified visually), and galaxy pairs very close to the final coalescence (separation $<10$ kpc).
\citet{robaina09} found a SFR enhancement of $\times 2$ in visual morphological mergers at redshifts $0.4<z<0.8$, which can increase to a maximum of $4$ if considering an Arp 220 SED template in the conversion between the luminosity at $24 \mu m$ and the total infrared luminosity for dusty starbursts.
At slightly higher redshifts ($1.9<z<2.1$), \citet{kaviraj13} found an enhancement of $\times 2.2$ in massive, visual morphological mergers.
More recently, \citet{shah22} have identified mergers and interacting systems in a sample of massive galaxies at $z_{spec}=0.5$-$3$. They consider mergers as single coalesced systems with clear morphological disturbances at visual inspection, while they define blended interactions as systems with at least two visually distinguishable galaxies but photometric measurements corresponding to their combined light, which have typical projected separations between $5$ and $10$ kpc.
While they find a significant SFR enhancement in mergers and interactions at $z<1.6$ by a factor of $\sim 2.5$, this effect almost vanishes in the higher redshift bin ($1.6<z<3$), where they measure weaker enhancements of $1.23$ and $1.06$, respectively, for mergers and blended interactions. 
Despite the variety of sample selection and merger identification methods adopted, these results suggest that in the early Universe the SFR enhancement induced by mergers is typically lower than at lower redshifts. 

\subsection{Burstiness of merging vs isolated galaxies}\label{sec:burstiness}

\begin{figure}[t!]
    \centering
    \includegraphics[angle=0,width=0.95\linewidth,trim={0.cm 3cm 19.5cm 0.cm},clip]{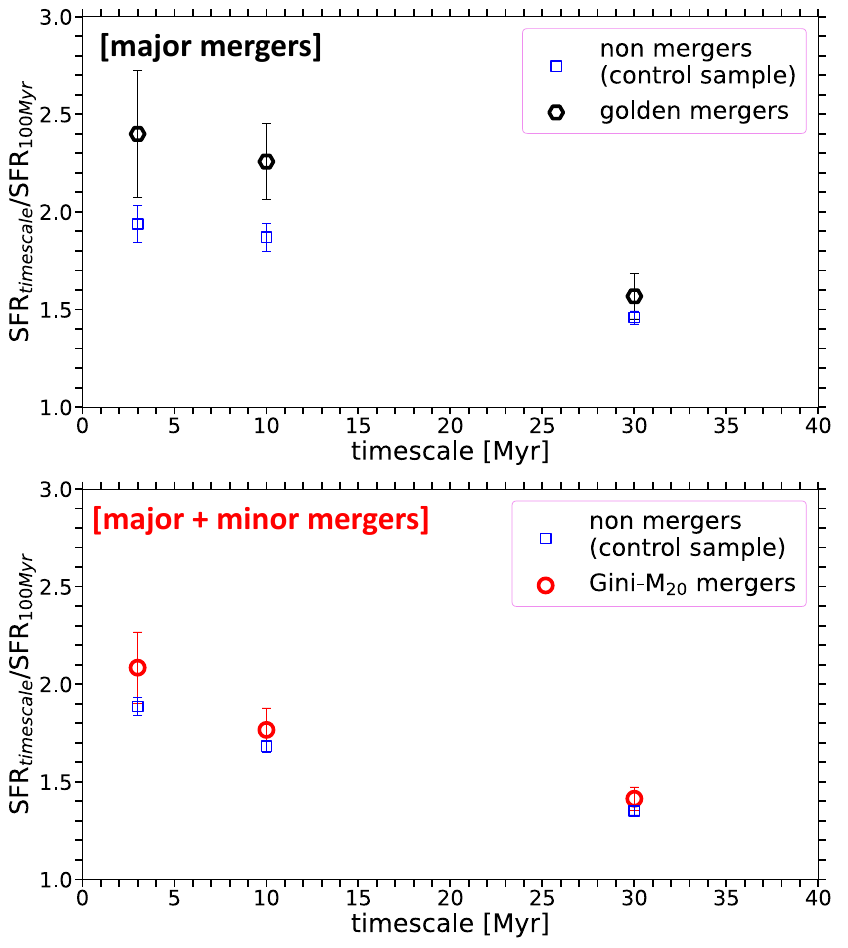}
    \vspace{-0.3cm}
    \caption{Median burstiness parameter (y axis), defined as SFR$_{\rm t-short}$/SFR$_{100 Myr}$, of the golden merger subset (upper panel) and Gini-M$_{20}$ merger subset (bottom panel) compared to the non-merger subset (respectively, the red circles and blue squares). The comparison is shown for three different SFR timescales `t-short' of $3$ Myr, $10$ Myr, and $30$ Myr (x axis). The red and blue error bars are the uncertainties on the median burstiness of the two subsets. 
    }\label{fig:burstiness}
    \vspace{-0.3cm}
\end{figure}

Thanks to our SED fitting approach, we can investigate the SFR enhancement in mergers over different timescales in a quantitative way. This can also inform us on the SFH of merging galaxies compared to those that are forming stars in a steady-state or by gas accretion. 
In particular, we calculate for each galaxy the burstiness parameter defined in Section \ref{sec:SED_fitting}, comparing the SFR averaged over the last $3$ Myr, $10$ Myr, and $30$ Myr, to the SFR averaged over $100$ Myr. 
To perform a statistical analysis, we estimate for each timescale the median burstiness of the golden and Gini-M$_{20}$ merger subsets, taking into account the entire stellar mass and redshift range of our analysis, and then 
compare these results to the median burstiness obtained for a mass and redshift matched control sample of isolated galaxies, following the same procedure described in Section \ref{sec:SFR_enhancement}. 

We find burstiness parameters ranging between $0.002$ and $300$ for the entire sample, with approximately gaussian distributions in logarithmic space, and median values of $2.2$, $1.7$, and $1.3$, respectively for the $3$ Myr, $10$ Myr, and $30$ Myr timescale (see Fig. \ref{fig:burstiness_histograms}). In addition, more than $\sim 75\%$ of the galaxies have a burstiness $>1$, with the first and third quartiles of the distribution being $\sim 1$ and $\sim 4$, respectively. This implies that the SFHs are rising on average for the stellar mass and redshift range considered in this study. %that on average the SFHs are rising in the last $10$-$1$ Myr. 
The $10$ Myr burstiness of the global population has a median value of $1.9 \pm 0.2$, and spans a similar range to those found by \citet{cole25} in a similar redshift and stellar mass range of our study, and using a similar SED fitting technique to our work. 

Focusing on the golden merger sample, it has median burstiness parameter of $1.57 \pm 0.12$ for the $30$ Myr timescale and $2.4 \pm 0.3$ for the $3$ Myr timescale (Fig. \ref{fig:burstiness}-\textit{top}). 
The isolated galaxy sample has median burstiness values $>1$, increasing toward shorter timescales, but mergers have a systematically higher burstiness than isolated galaxies for all the three timescales. 
In the shortest timescale cases ($<10$ Myr) the burstiness in major mergers is a factor of $1.23 \pm 0.12$ higher than in isolated galaxies. For the $30$ Myr timescale, the difference between mergers and non mergers becomes not significant (burstiness ratio $= 1.07 \pm 0.08$).  
This suggests that the impact of major mergers on the galaxy's SFRs mainly occurs in the last $10$ Myr from the time of observation, during which they induce stronger bursts of star-formation compared to normal, isolated galaxies at the same cosmic epoch and stellar mass. This result can be naturally explained by our morphological merger selection, which identifies more advanced merger phases (e.g., compared to the close pair selection), which are likely closer to the final coalescence. The enhanced burstiness also reflects the bursty nature of star-formation during the merger episode.

Finally, we explore the burstiness parameter of the Gini-M$_{20}$ sample. In this case, the burstiness is generally lower than in the previous case, increasing from a median of $1.41 \pm 0.06$ for the $30$ Myr timescale to $2.0 \pm 0.2$ for the $3$ Myr timescale (Fig. \ref{fig:burstiness}-\textit{bottom}). We obtain marginally significant burstiness enhancements (at $\sim 1 \sigma$ significance) of $1.11 \pm 0.10$, $1.05 \pm 0.07$, and $1.05 \pm 0.05$, respectively for the $3$ Myr, $10$ Myr, and $30$ Myr timescales. 
This implies that major mergers likely have a higher effect on the burstiness than minor mergers, which is also consistent with the stronger enhancement of their SFRs at fixed stellar mass and redshift (Fig. \ref{fig:SFR_mass}).

\section{Discussion}\label{sec:discussion}

\subsection{Comparison to predictions of semi-analytic models and simulations}\label{SAM_comparison}

To understand our findings on the merger induced SFR enhancement across cosmic time, it is useful to compare them to the predictions of semi-analytic models of galaxy formation, and to hydro-dynamical simulations of galaxy mergers at different redshifts.
First, we consider the Semi-analytic Forecasts for JWST simulation catalogue. This project makes predictions for the distribution and properties of galaxies from $z=0$ to redshift $10$, and is based on the Santa Cruz semi-analytic model (hereafter SC-SAM\footnote{The catalog and relevant information can be found at this webpage \url{https://flathub.flatironinstitute.org/sam-jwst-deep}}) of galaxy formation \citep{somerville15,somerville21}. As shown in \citet{yung19a} and \citet{yung19b}, the physical framework and implementation of this model can reproduce the observed photometry of galaxies from $z=0$ to $z=10$, and the distribution functions of fundamental galaxy properties (e.g., M$_{\rm UV}$, M$_\ast$, SFR) and related scaling relations, down to the lowest masses of M$_\ast \sim 10^7$ M$_\odot$, which can now be reached with JWST at the EoR. More details on the physics considered in these models can be found in a series of studies \citep{yung19a,yung19b,yung20,yung22}.

For this comparison, we consider the JWST ultradeep lightcones, which are a set of $8$ deep-field lightcone realizations, each covering an area of $132$ arcmin$^2$. This includes a total number of $\sim 2$ Million simulated galaxies from $z=0$ to $z=10$, which provides enough statistics for our work while not excessively overloading the computational effort required for the subsequent analysis. In this simulation, the lightcone contains galaxies well below those observable with JWST.  
For our purposes, we apply a magnitude cut of $30$ in the F444W band, which is approximately the depth of JWST imaging observations used in this study, and a redshift selection ranging from $5$ to $10$ to match our selected galaxy sample. 
For each galaxy this catalog includes the information on the time since it last went through a major or minor merger, as two separate columns $t_{\rm major merge}$ and $t_{\rm merge}$. It also includes information on the stellar mass of the primary galaxy that went through a merger and on the instantaneous SFR, which is the quantity more directly comparable to our SFR values. 
In this work, we identify minor and major mergers in the models by selecting galaxies that underwent, respectively, a minor or major merger (down to mass ratios of $10$:$1$ and $4$:$1$) within the observability timescales defined in Section \ref{sec:merger_timescales} for the Gini-M$_{20}$ and the golden merger criteria (the function represented by the red curve in Fig. \ref{fig:merger_timescales}).  

\begin{figure}[t!]
    \centering
    \includegraphics[angle=0,width=1\linewidth,trim={0.cm 0.cm 0.cm 0.cm},clip]{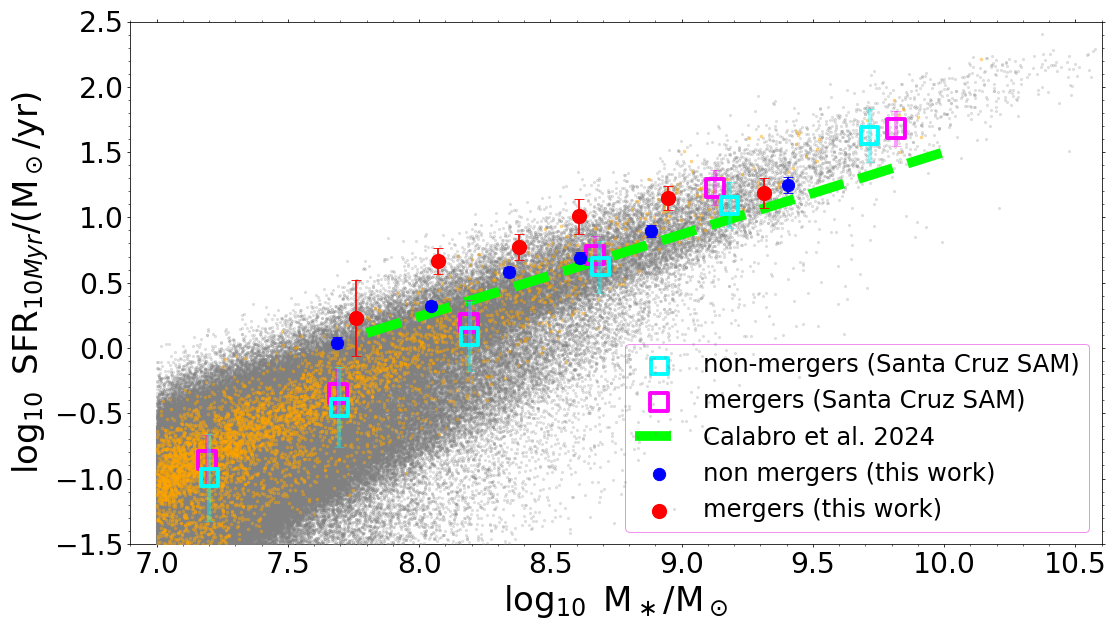}
    \includegraphics[angle=0,width=1\linewidth,trim={0.cm 0.cm 0.cm 0.cm},clip]{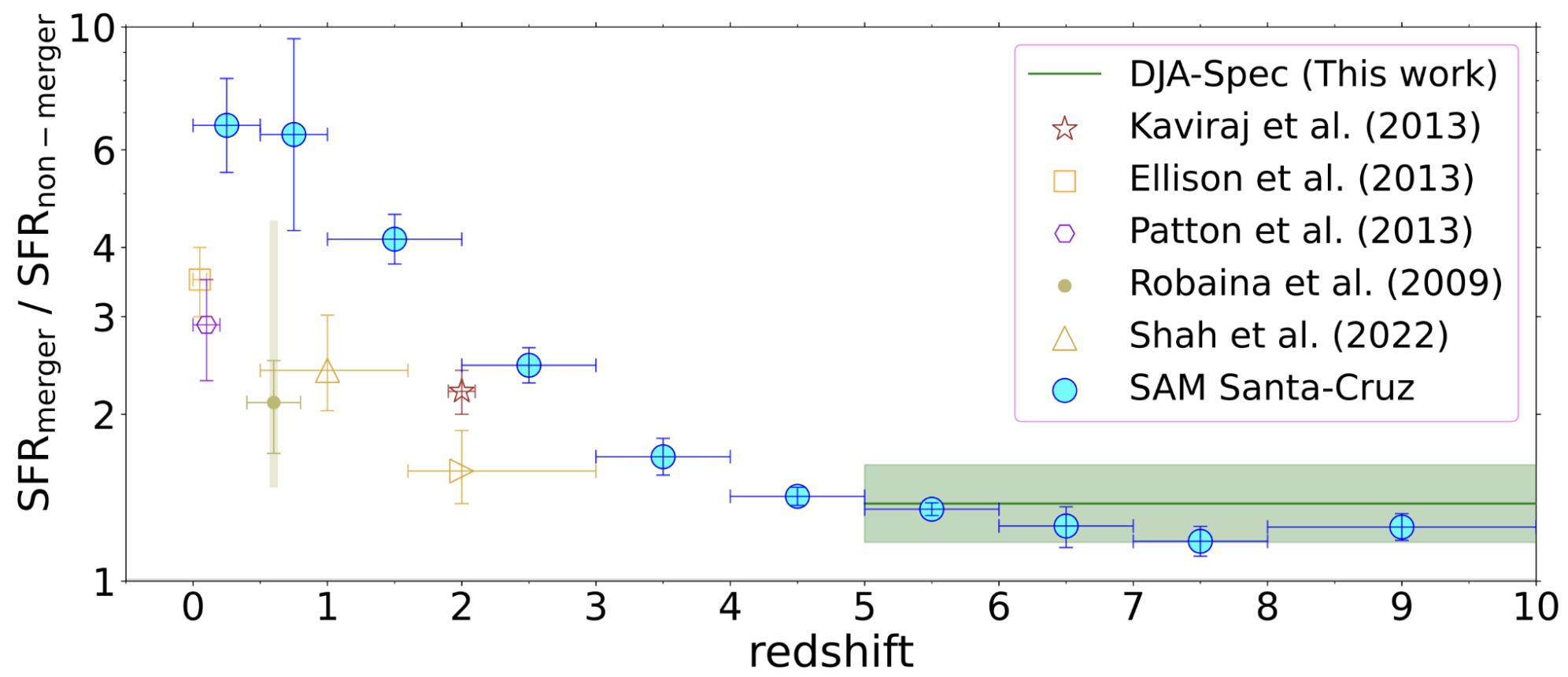}
    \vspace{-0.5cm}
    \caption{Predictions of the Santa Cruz SAM. \textit{Top:} SFR - M$_\star$ diagram for mergers (orange small circles) and non-mergers (gray small circles). The median SFRs in M$_\star$ bins are shown with fuchsia big empty squares and cyan empty squares, respectively for the merger and non-merger subsets, and the error bars represent the median absolute deviation (MAD) of the SFRs in the same bin. The observed SFR - M$_\star$ relation is shown with blue and red filled circles, respectively, for the golden merger sample and for non mergers. \textit{Bottom:} The median SFR enhancement of the merger vs the non-merger sample (SFR$_{\rm med, merger}$/SFR$_{\rm med, non-merger}$) from $z=0$ to $z=10$, as predicted by the Santa Cruz SAM. The range of SFR enhancement observed in this work is highlighted with a green shaded rectangle. This is compared to the SFR enhancement found in mergers at lower redshifts \citep{robaina09,kaviraj13,patton13,ellison13,shah22}, which reach a maximum factor of $\sim 4$.
    }\label{fig:SFR_enhancement_SantaCruz}
    \vspace{-0.2cm}
\end{figure}

There are several approximations in this procedure. 
First, we assume that the number of galaxies that have experienced a merger in the last few Myr is the same as the number of galaxies that are going to coalesce within the same time, which are those identified by our morphological selection.  
In addition, the catalog associated to the simulated lightcone reports mass ratios that are based on the total baryonic mass of the galaxies (thus including gas, dust, and stars), while in the observations we consider the stellar mass ratios of colliding galaxies when defining the merger type and the corresponding observability timescale. 

We now investigate the SFR - stellar mass diagram of merging and non-merging galaxies in the SC-SAM, as done for the observational data in Fig. \ref{fig:SFR_mass}. 
We compute the median SFR in $7$ increasing stellar mass bins from $10^7$ to $10^{10.5}$ M$_\odot$. The results for major mergers are shown in Fig. \ref{fig:SFR_enhancement_SantaCruz}-\textit{top}. They indicate that the merger subset has systematically higher SFRs compared to the non-merger subset in all the stellar mass bins, with a significant enhancement by more than $10 \sigma$ in all except the last bin, and a median SFR enhancement of $0.11 \pm 0.02$ dex (a factor of $\sim 1.3$) over the whole mass and redshift range, comparable to that found in this work from observations. A similar result is obtained from the models when we also include minor mergers in the analysis. 

To understand the lower SFR enhancement of high redshift ($z>5$) mergers compared to low redshift ones, as seen in Section \ref{sec:SFR_enhancement}, we analyze the redshift evolution of the SFR enhancement predicted by the models. We calculate the median SFR enhancement of the merger subset in different redshift bins, from $0$ to $10$, adopting a uniform cut on the stellar mass as $\log_{10}$ M$_\ast$/M$_\odot$ $\geq 8$. 
In Fig. \ref{fig:SFR_enhancement_SantaCruz}-\textit{bottom}, we can see that the maximum SFR enhancement (by a factor of $6_{-2}^{+3}$) occurs in low redshift mergers ($z \leq 1$), while it strongly decreases by $\sim 3.5$ times in earlier mergers, between $z=1$ and $z=4$. At $z \gtrsim 6$, the enhancement stabilizes around a factor of $1.2$. 
This analysis shows that the SC-SAM predicts a strong evolution of the SFR enhancement in mergers from the local Universe to the EoR, consistent with the trend inferred from observations. 
Most importantly, the models corroborate our results indicating that mergers have an impact on the star-formation of galaxies back to the earliest cosmic epochs that we have explored.
We also note that the models predict a maximum SFR enhancement that is significantly higher than those obtained in observations (see Section \ref{sec:SFR_enhancement}). 
This difference might be in part due to the observed SFRs calculated over timescales of $\sim 100$ Myr, while those from the models are instantaneous. Moreover, the merger duration at low redshifts significantly increases, hence the instantaneous SFRs that we use for comparison are no longer representative of the entire merger process, thus they should be interpreted as the peak SFR enhancement that can last for short periods of time. 

The role of mergers in the star-formation activity of high redshift galaxies is also predicted by hydrodynamical simulations. In particular, \citet{fensch17} performed pc-resolution numerical simulations of mergers in the low gas fraction ($= 10 \%$) and high gas fraction case ($= 60 \%$), for different orbits (rotation and impact parameter) of the interacting galaxies. These gas fractions are typical, respectively, of $z=0$ and $z=2$ galaxies. They found that the SFR enhancement induced by the merger in the high f$_{\rm gas}$ case ranges between a factor of $1.5$ and $3$ (depending on the rotation of the galaxies and the impact geometry), and is approximately $10$ times lower than in the low f$_{\rm gas}$ case.
More recently, \citet{schechter25} analyzed mergers in the Illustris TNG-50 simulation. At $z<0.5$, they find that the SFR in mergers can be enhanced by a factor of $\sim 2$ close to the final coalescence. This effect is less pronounced at higher redshifts, stabilyzing around an enhancement factor of $1.5$ for major mergers at $1<z<3$ (the highest redshift bin of their analysis) and in the stellar mass range between $10^8$ and $10^9$ M$_\odot$. This effect is comparable to that found at $z>5$ in this study (average SFR enhancement factor $=1.4 \pm 0.1$), considering galaxies in the same stellar mass range $10^8$-$10^9$ M$_\odot$. \citet{schechter25} also find that a SFR enhancement at $z=1$-$3$ is also present in minor mergers, although it is slightly lower (by $5\%$-$10\%$) than in major mergers, in agreement with our results. 

The above simulations help us to understand the physical differences between local mergers and the high redshift ones, giving an explanation of the observed results.  
In the local Universe, the SFR increase in mergers is caused by several factors, including enhanced gas inflows toward the galactic center (typically in the inner $1$kpc), increased gas turbulence, and compressive tides induced by gravitational torques during the interaction \citep{renaud08,renaud09,sparre22}. At higher redshifts the above mechanisms are already efficient in isolated systems due to the higher f$_{\rm gas}$ of typical star-forming galaxies, as we reach f$_{\rm gas}$ $\sim 50 \%$ at $z \simeq 2$ from $\sim 10 \%$ in the local Universe \citep{daddi10,tacconi10}. Cosmological gas inflows and violent disk instabilities represent competing channels to mergers for the increase of gas turbulence, inward migration of gas, and SFR increase in these galaxies \citep{bournaud12,gabor14}. For example, the relative increase in baryonic mass in the central $1$kpc due to merger induced inflows is approximately half the value of that in the local Universe (\Citealt{fensch17} and references therein). 
Another physical mechanism to consider is that at higher redshifts the fraction of galaxies with a central bulge or with a bar decreases \citep{sheth08,fragkoudi20,costantin25}. Therefore, the average star-forming galaxy population lacks the additional torque induced by these structures, which strongly contributes to the infall of gas in low redshift mergers.

\subsection{The merger rate and the contribution of mergers to the galaxy mass assembly}\label{mass_assembly}

In this section we explore the redshift evolution of the merger rates of galaxies $\mathcal{R}_\mathrm{M}$(z, M$_\star$), defined as the number of merger events per galaxy per unit time following a similar formalism to \citet{conselice22} and \citet{puskas25_mergers} as : 
\begin{equation}
    \mathcal{R}_\mathrm{M}(z, M_\star) = \frac{f_\mathrm{merge}(z, M_\star)}{\tau_\mathrm{merge}(z)} = \frac{f_\mathrm{tracer}(z, M_\star) \times C_\mathrm{merg, tracer}}{\tau_\mathrm{obs, tracer}(z) } ~({\rm Gyr^{-1}}),
    \label{Eq:merger_rate1}
\end{equation}
where $f_\mathrm{merge}$ is the merger fraction for a specific redshift $z$ and stellar mass M$_\star$, $\tau_\mathrm{merge}$ is the dynamical time of the merger as defined in Section \ref{sec:merger_timescales}, $f_\mathrm{tracer}$ is the observed fraction of mergers for a specific merger tracer as derived in Section \ref{sec:merger_fractions}, $C_\mathrm{merge, tracer}$ is the fraction of observed galaxy pairs that will actually merge in the future, and finally $\tau_\mathrm{obs, tracer}$(z) is the visibility timescale of the tracer, as introduced in Section \ref{sec:merger_timescales}. In the following part of the paper, we simply use the term $\mathcal{R}_\mathrm{M}$(z), which refers to the merger rate $\mathcal{R}_\mathrm{M}$(z, M$_\star$) of galaxies with a similar mass of our sample (M$_\ast$ $\sim 10^7$-$10^9$ M$_\odot$).
We adopt $C_\mathrm{merg, tracer}=1$ in this study, as we assume that all the mergers identified through strong morphological disturbances will actually coalesce in the future. 
However, this quantity may be substantially lower, typically in the range $0.4$-$1$ for the photometric pair method \citep{oleary21,puskas25_mergers,duan25}, indicating that the observed pair has a probability of $40 \%$-$100 \%$ to actually coalesce into a single galaxy at later epochs, due to uncertain redshifts, unknown relative velocities, and chance projection effects. However, the results obtained by the most recent photometric pair studies are based on the assumption that $C_\mathrm{merg, tracer}=1$ \citep[e.g.][]{conselice22,duan25,puskas25_mergers}.

\begin{figure}[t!]
    \centering
    \includegraphics[angle=0,width=\linewidth,trim={0.cm 2cm 19.cm 0.cm},clip]{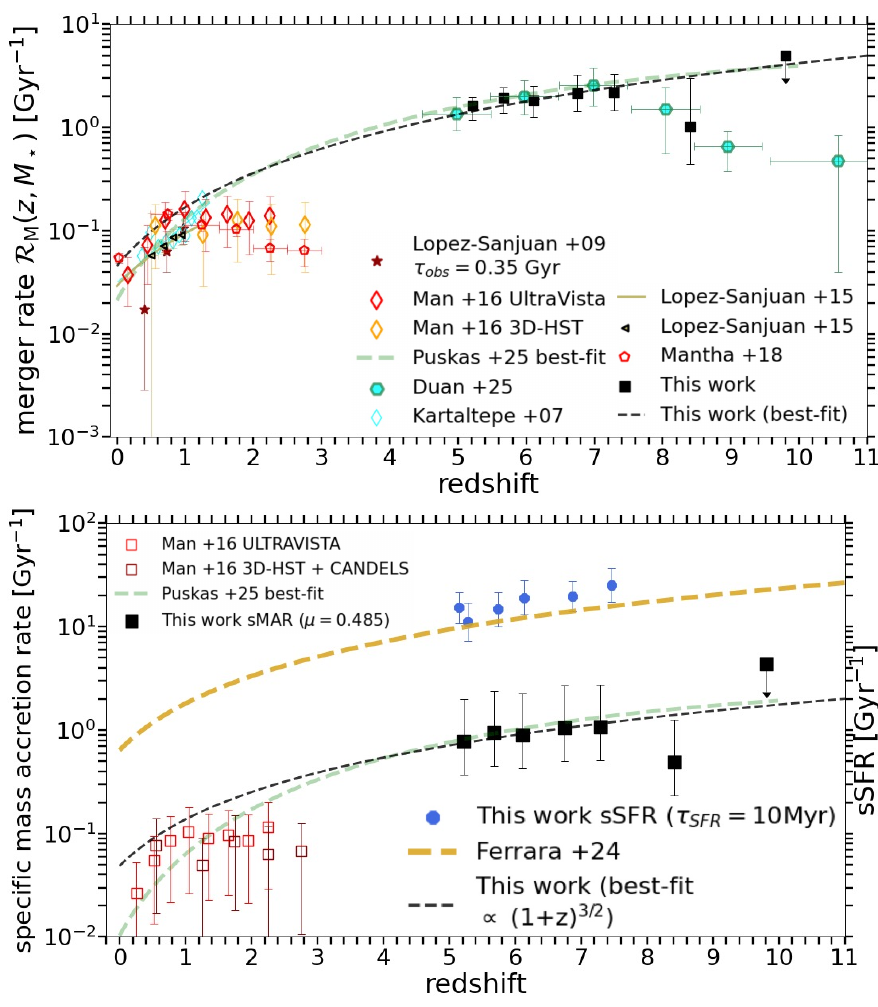}
    \vspace{-0.6cm}
    \caption{\textit{Top:} Major merger rate derived in this study using the golden merger sample defined in Eq. \ref{eq:merger_criterion} (black points), compared to other results from the literature at similar redshifts \citep{duan25,puskas25_mergers} and at lower redshifts \citep{kartaltepe07,lopez-sanjuan09,lopez-sanjuan15,man16,mantha18}. Our error bars incorporate the uncertainty due to the mass ratio $\mu$ ranging between $1$ and $0.25$, typical of major mergers. The best-fit to all observed data from $z=0$ to $z=10$ is shown with a black dashed line. 
    \textit{Bottom:} Major merger sMAR obtained in this study (black points), and in previous works at lower redshifts (red empty squares), including \citet{man16}. The best-fit relation considering our observed points and low redshift results in the functional form $\mathcal{R}_M=a \times (1+z)^b$ is shown with a black dashed line, while the one obtained by \citet{puskas25_mergers} with the photometric pair method is drawn with a green dashed line. This function is compared to the redshift evolution of the sSFR from this work (blue circles) and from the model of \citet{ferrara24} (golden dashed line). 
    }\label{fig:merger_rate_golden_sample}
    \vspace{-0.4cm}
\end{figure}

\begin{figure}[t!]
    \centering
    \includegraphics[angle=0,width=\linewidth,trim={0.cm 2cm 19.cm 0.cm},clip]{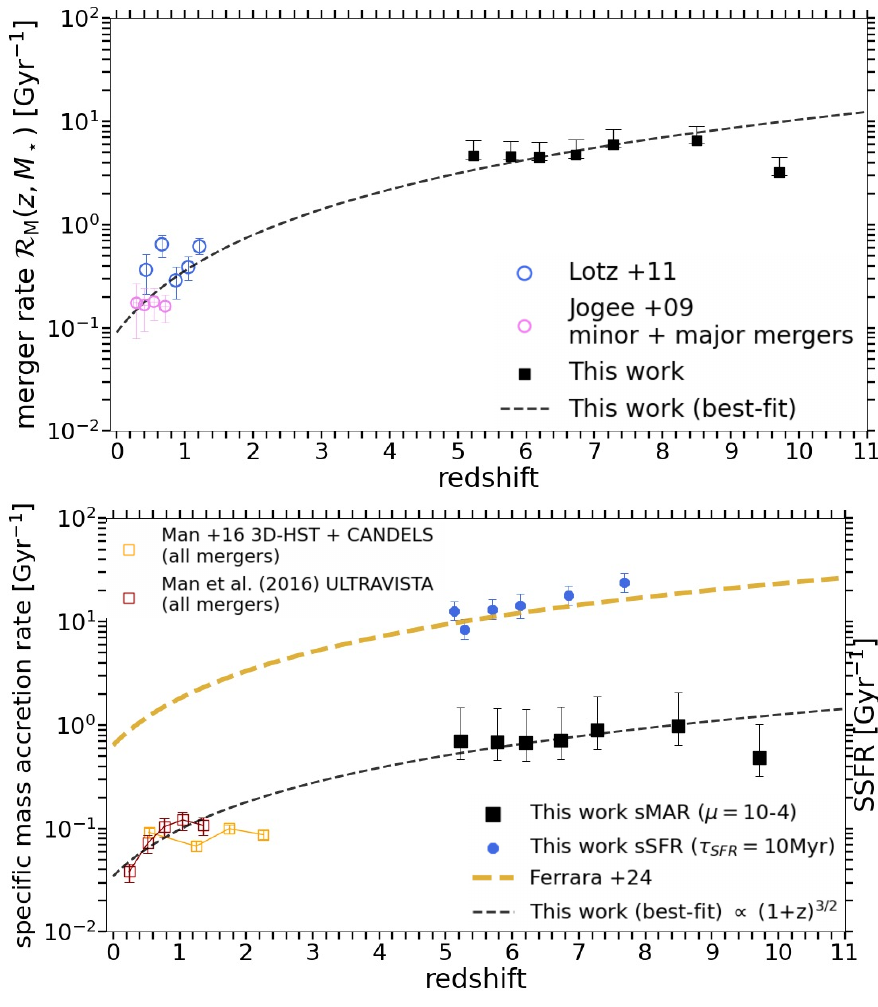}
    \vspace{-0.6cm}
    \caption{Same as in Fig. \ref{fig:merger_rate_golden_sample}, but using the Gini-M$_{20}$ merger sample to calculate minor+major merger rates.
    Our findings are compared to those obtained at lower redshifts by \citet{lotz11,jogee09,man16}.
    }\label{fig:merger_rate_ginim20}
    \vspace{-0.4cm}
\end{figure}

The merger rate $\mathcal{R}_\mathrm{M}$(z) can be calculated separately for major and minor+major mergers. 
To estimate the major merger rate from Eq. \ref{Eq:merger_rate1}, we use the fration of golden mergers f$_{m,golden}(z)$ estimated in Section \ref{sec:merger_fractions}, because it is more stable against variations of the gas fraction and mass ratio $\mu$ compared to the fraction of asymmetry mergers. The results are shown in the top panel of Fig. \ref{fig:merger_rate_golden_sample},where we can see that the major merger rate slightly increases from redshift $5$ to $7.5$, ranging between $1$ and $3$ merger events per galaxy per Gyr. At later epochs ($z > 7.5$), $\mathcal{R}_\mathrm{M}(z)$ drops again to $1$ major merger per galaxy per Gyr (Fig. \ref{fig:merger_rate_golden_sample}-top). However, given the larger uncertainty, it would still be consistent with the average increasing trend obtained from the lower redshift bins. Considering the entire redshift range, this method yields an average major merger rate $\mathcal{R}_\mathrm{M}$ of $1.8 \pm 0.3$ mergers/galaxy/Gyr.

If we compare this result to lower redshift studies, we can notice that $\mathcal{R}_\mathrm{M}(z)$ has a strong evolution from $z=0$ to $z=10$ by more than one order of magnitude, going from $\leq 0.1$ merger/galaxy/Gyr at $z \leq 1$ to $\geq 1$ merger/galaxy/Gyr at $z>5$ (Fig. \ref{fig:merger_rate_golden_sample}-top). 
Considering our results and low redshift ($z<1$) data points \citep{lopez-sanjuan09,lopez-sanjuan15,mantha18,kartaltepe07,man16}, we can fit the major merger rate evolution with a similar function to that adopted in \citet{conselice22} as:
\begin{equation}\label{Eq:merger_rate_evolution}
 \mathcal{R}_\mathrm{M}(z) = a \times (1+z)^b ,
\end{equation}
with $a_{major-M}=0.045 \pm 0.009$ and $b_{major-M}= 1.9 \pm 0.1$. 
The major merger rate evolution is remarkably consistent with the relation obtained by \citet{puskas25_mergers} with the photometric pair method across all $\sim 12$ Gyr of cosmic time for a similar stellar mass range of our study. 
At $z>5$, our results are also consistent to those found with the close pair method by \citet{duan25} at similar stellar masses. In this case, to ensure a fair comparison, we have considered the pair fractions f$_{\rm obs,pairs}$ obtained by \citet{duan25} for maximum pair separations of $30$ kpc (see their Figure~5-top), and we calculated the merger fraction with our Eq. \ref{Eq:merger_rate1}, assuming $C_\mathrm{merg} = 1$, and $\tau_{\rm obs,pairs}$ given by the expression in Eq. \ref{Eq:merger_timescale_conselice22} with $\bar{\mu} = 0.485$ valid for major mergers \citet{puskas25_mergers}. However, also using f$_{\rm obs,pairs}$ with their fiducial setting (i.e., pair separations of $20$ to $50$ kpc), we would find major merger rates in agreement with our study.  

To evaluate the minor+major merger rate, we use instead the fraction of Gini-M$_{20}$ mergers f$_{m,Gini-M_{20}}$ derived in Section \ref{sec:merger_fractions}, as the criterion Gini$+0.14\times$ M$_{20}>0.33$ is sensitive to mergers down to $\mu=0.1$ \citep{lotz11}.  As shown in Fig. \ref{fig:merger_rate_ginim20}-top, the minor+major merger rate follows a similar redshift evolution as the major merger rate but with systematically higher values, going from $\sim$4 mergers/galaxy/Gyr at z$\simeq$5 to $\sim$7 mergers/galaxy/Gyr at z$\simeq$7, before dropping at $\sim$3 mergers/galaxy/Gyr in the last redshift bin at z$>9$ (Fig. \ref{fig:merger_rate_ginim20}-top). The average $\mathcal{R}_\mathrm{M}$ across the entire range $5<z<10$ is $5.0 \pm 0.3$ mergers/galaxy/Gyr, that is, approximately $\times 3$ higher compared to major mergers alone. The merger rate values estimated in different redshift bins are reported in Table \ref{table:results_merger_rates_sMAR}.
Using Equation \ref{Eq:merger_rate_evolution} to fit our data in combination with low redshift ($z<1.5$) studies \citep{lotz11,jogee09}, we obtain $a_{all-M}= 0.09 \pm 0.01$ and $b_{all-M}= 2.0 \pm 0.1$. This again indicates a strong evolution of the major+minor merger rate across $\sim 12$ Gyr of cosmic history, with mergers becoming more numerous and more frequent in the early Universe.   
We note that, even though the major and major+minor merger rates at $z>5$ are much higher compared to the local Universe, the amount of mass added per merger is presumably lower, due to the typical smaller masses of high redshift galaxies compared to the samples analyzed at lower $z$.

We can finally estimate the contribution of major and minor+major mergers to the stellar mass growth of galaxies across different epochs. 
In particular, we calculate the specific mass accretion rate (sMAR) of mergers using the formalism of \citet{duncan19} and \citet{puskas25_mergers}, as :
\begin{equation}\label{Eq:specific_mass_accretion_rate}
sMAR(z) = \mathcal{R}_\mathrm{M}(z) \times \bar{\mu},
\end{equation}
where $\mathcal{R}_\mathrm{M}(z)$ is defined in Equation \ref{Eq:merger_rate_evolution} and $\bar{\mu}$ is the average mass ratio of the sample. Since we are probing more advanced merger phases compared to the close pair method, and given that in general the individual merging galaxies cannot be clearly identified, it is not feasible to assess statistically the original mass ratio of the colliding galaxies in our sample. We thus consider the value calculated for galaxy pairs in photometric samples at the same redshift and stellar mass range of this study.
In particular, for golden mergers, we consider the value $\bar{\mu} = 0.485$ \citep{puskas25_mergers}. The associated error bars are derived by assuming that $\bar{\mu}$ can vary in a range from $0.25$ to $1$, consistent with the major merger definition. 
For the Gini-M$_{20}$ merger sample, as they trace also minor mergers, we consider a $\bar{\mu} = 0.15$ \citep{lotz11}, with error bars incorporating a $\bar{\mu}$ variation between $0.1$ and $1$. 

With these assumptions, we calculate the sMAR of our merger sample in different redshift bins, which are reported in Table \ref{table:results_merger_rates_sMAR}.
We find that, in the redshift range of our analysis, the sMAR of major mergers varies between $0.5$ and $1$ Gyr$^{-1}$, with a median of $0.92 \pm 0.08$ (Fig. \ref{fig:merger_rate_golden_sample}-bottom). As for $\mathcal{R}_\mathrm{M}(z)$, the lower sMAR found for major mergers in the highest redshift bin has a higher uncertainty and is still consistent with the global increasing trend derived at lower redshifts. 
For the minor+major merger sMAR, while comprised in the same range as major mergers, has a lower median of $0.71 \pm 0.06$ (Fig. \ref{fig:merger_rate_ginim20}-bottom). 

To assess the relative importance of mergers in galaxy growth, we can compare the major and minor+major merger sMAR to the specific mass accretion due to star-formation, which is by definition the sSFR.  
For our sample we use the SFRs averaged over a timescale of $10$ Myr, which yield sSFR(z) that in reasonable agreement with the theoretical evolution derived in \citet{ferrara24} as $0.64 \times (1+z)^{1.5}$.
We show the results in Figures \ref{fig:merger_rate_golden_sample} and \ref{fig:merger_rate_ginim20} for major and minor+major mergers, respectively. We find that at redshifts $5<z<10$ both sMAR are approximately an order of magnitude smaller than the sSFR.  
Dividing the two quantities, the merger contribution to the mass assembly of galaxies varies between $5\%$ and $10\%$, much lower than the contribution from star-formation in isolated galaxies. 

If we also fit the sMAR derived at $z<3$ by \citet{man16}, we find that the sMAR(z) follows an evolution similar to sSFR(z), and can be well fit by a power-law function with the same exponent of $1.5$ but a lower normalization. Writing the evolution as $c \times (1+z)^{1.5}$, we find $c= 0.048 \pm 0.002$ for the major merger sample and $c = 0.052 \pm 0.003$ for the minor+major merger sample. % However, again this function does not reproduce the drop of merger rate and sMAR observed in the last redshift bin mentioned above. 
Therefore, all merger types provide a lower contribution to the build up of galaxies compared to secular star-formation at all cosmic epochs, and do not play a bigger role in the early Universe. This is consistent with the conclusions of \citet{puskas25_mergers}, whose sMAR(z) for major mergers is in agreement with our results (Fig. \ref{fig:merger_rate_golden_sample}-\textit{bottom}). 
We finally note that assuming a longer observability timescale (up to $100$ Myr) for our morphological merger criteria, the merger rates and the sMAR would be $\sim 0.47$ dex lower, strenghtening our conclusions on the minor role of mergers in galaxy mass assembly.

\section{Summary and conclusions}

We have investigated the merger fraction at $z>5$ in a spectroscopically selected sample of galaxies, with mergers identified based on rest-frame optical morphological disturbances (traced by the Gini, M$_{20}$, and asymmetry parameters) in F444W NIRCam images. This provides a complementary approach, tracing shorter merger phases, to the close photometric pairs adopted in similar previous studies. We have also explored the role of mergers in the star-formation rate enhancement in the early Universe, and their contribution to the galaxy mass build-up across different epochs. 
We summarize our findings as follows :
\begin{itemize}
\item the morphological merger fraction does not strongly evolve with redshift, remaining approximately constant from $z\simeq1$ to the redshifts explored in this study ($5<z<10$). % , and eventually decreasing only at $z>8$.   
\item the absolute value of the merger fraction depends on the type of merger and on the merger observability timescale traced by each morphological parameter, increasing for minor mergers and for longer merger observability. In particular, it varies from $\sim 5 \%$ for the gold merger sample (which traces major mergers with $\tau_{obs,golden}=$ $30$-$10$ Myr between $z=5$ and $10$), to $\sim 8 \%$ for Asymmetry mergers having longer observability timescales by approximately a factor of $2$. % of $\simeq 400$ Myr at $z=1$.  XXXX Forse scrivilo diversamente XXX We are interested in the timescale at $z>5$, not at $z=1$. XXXX
The merger fractions further increase to $\sim 13 \%$ when considering Gini-M$_{20}$ mergers, which likely trace minor+major mergers (down to a mass ratio $\mu = 0.1$) with $\tau_{obs, z=1}\simeq 200$ Myr. 
\item mergers at redshifts $5<z<10$ have a significant impact (although much lower than at $z<1$) on the SFR of galaxies, with an average SFR enhancement by a factor of $\sim 1.8$ compared to isolated star-forming galaxies at the same redshift and stellar mass. The enhancement is higher for shorter timescale SFRs and for morphological indicators with shorter merger observability timescales, in agreement with a short-lived, bursty nature of merger triggered star-formation. 
\item after accounting for the observability timescales of each merger tracer, we find major merger rates of $\sim 2$ merger/galaxy/Gyr on average at redshifts $5<z<10$, and $\sim 5$ merger/galaxy/Gyr for minor+major mergers. Combined with low redshift results, the merger rates show a strong evolution in redshift, increasing approximately by one order of magnitude from $z = 1$ to $z = 7$. This might be in part due to the intrinsically shorter merger duration in the early Universe. The derived major merger rate is in agreement with that inferred using the close pair method on photometric galaxy samples in a similar mass and redshift range of this study. Despite the higher merger rate, the amount of mass added per merger at $z>5$ is likely lower than at $z<1$, due to the typical smaller masses of high-$z$ galaxies compared to those probed at lower $z$.
\item we find that mergers with a mass ratio $\mu > 0.1$ contribute $\sim 5$-$10 \%$ to the mass build-up of galaxies at $5<z<10$ compared to in-situ star-formation, suggesting that mergers are not the dominant process in galaxy growth as in the low redshift Universe.
\end{itemize}  

This study sets the groundbase for a more detailed investigation of the effects of mergers on other galaxy properties, including dust attenuation, metallicity, ionization, and ionizing continuum escape, through the use of spectroscopic data. In the future, this work will benefit from the growing number of spatially resolved spectroscopic observations, which will allow us to explore where most of the star-formation occurs during merger events.
Moreover, future imaging observations with the Extremely Large Telescope (ELT), thanks to its greater sensitivity and resolution compared to JWST, will enable an increase in the statistics of mergers in the early Universe and in low-mass regime (M$_\ast$ $<10^7$ M$_\odot$), and the identification of late-stage mergers with very close nuclei or post-merger systems with faint residual interaction features.

\begin{acknowledgements}
We thank the referee for carefully reading the paper and providing valuable comments and suggestions. The research activities described in this paper were carried out with contribution of the Next Generation EU funds within the National Recovery and Resilience Plan (PNRR), Mission 4 – Education and Research, Component 2 – From Research to Business (M4C2), Investment Line 3.1 – Strengthening and creation of Research Infrastructures, Project IR0000034–“STILES – Strengthening the Italian Leadership in ELT and SKA”. 
\end{acknowledgements}

{}

\clearpage

\appendix

\section{Definition of morphological parameters}\label{appendix1:definition_of_morphological_parameters}

We briefly present here the main definition of the three morphological parameters adopted in this study (Gini, M$_{20}$, and asymmetry), while referring to previous papers for further details.
The Gini coefficient is defined as : 
\begin{equation}
    G=\frac{1}{\bar{X} n(n-1)} \sum_{i}^{n}(2 i-n-1) X_{i}
\end{equation}
where $X_i$ is the intensity of the $i^\mathrm{th}$ pixel, $n$ is the total number of pixels of the galaxy, and $\bar{X}$ is the mean intensity. The Gini parameter varies between $0$ and $1$ by construction, distinguishing between galaxies with more homogeneous %evenly distributed 
flux distributions (low Gini) and more unequal flux distributions (high Gini) % $>0.5$)
, which is more typical of mergers. 
In all cases, the pixels are assigned to the galaxy based on the segmentation map derived through the photutils package following \citet{treu23}, and adopting a flux threshold for detection of $3\sigma$ above the background.

The second-order moment of brightness $M_{20}$, is defined as : 
\begin{equation}
    M_{20}=\log _{10}\left(\frac{\sum_{i} M_{i}}{M_{tot}}\right), \text { with } \sum_{i} f_{i}<0.2 f_{\text{tot}} \\
\end{equation}
\citep{lotz04}, where $f_{\text{tot}}$ is the total flux of the galaxy pixels, $M_i$ is the second-order moment of brightness of each pixel, and $M_{\text{tot}}=\sum_{i}^{n} M_{i}=\sum_{i}^{n} f_{i}\left[\left(x_{i}-x_{c}\right)^{2}+\left(y_{i}-y_{c}\right)^{2}\right]$, where $f_i$ is the pixel intensity, $x_{i}$ and $y_{i}$ are the pixel coordinates, and $x_{c}$ and $y_{c}$ represent the galaxy center where $M_{\text {tot }}$ is minimized. The sum is done over the brightest $20\%$ pixels of the galaxy identified through the segmentation map. $M_{20}$ typically varies between $-3$ and $0$. Smooth, normal galaxies have lower $M_{20}$, while the presence of bright off-centered structures, such as multiple nuclei and tidal tails, tend to increase $M_{20}$. As a result, merger systems typically have higher values of this parameter. 

Lastly, the asymmetry ($A$) is defined as:
\begin{equation}
    A=\frac{\Sigma\left|I -I_{\pi}\right|}{\Sigma I}-A_{bkg}
\end{equation}
\citep{conselice00} where $I$ is the original cutout image, $I_{\pi}$ is the image rotated by 180 degrees, and $A_{\text{bkg}}$ is the asymmetry of the background.  Also this quantity varies between $0$ and $1$, with merging galaxies having higher values as due to their asymmetric and more irregular structure compared to non-mergers.

\section{Cutouts of galaxies}\label{appendix:cutouts}

\begin{figure}[t!]
    \centering
    \includegraphics[angle=0,width=\linewidth,trim={0.cm 0.cm 7.cm 0.cm},clip]{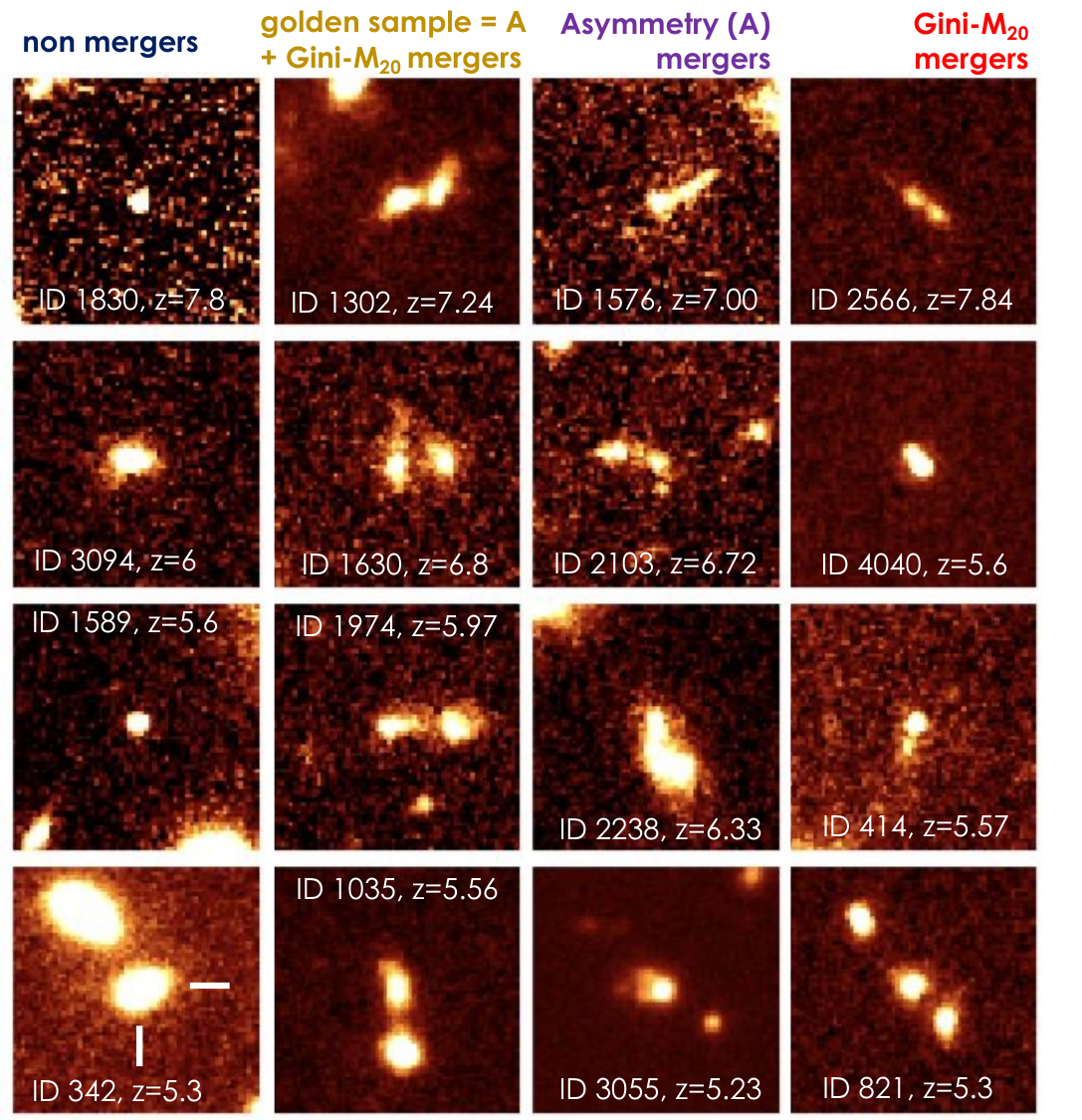}
    \vspace{-0.1cm}
    \caption{$3\arcsec \times 3\arcsec$ F444W cutouts of example galaxies selected in this work and belonging to different morphological classes, as defined in Section \ref{sec:merger_identification}. The columns include, from left to right, non-merger galaxies, gold mergers, asymmetry mergers, and Gini-M$_{20}$ mergers, representing each quadrant of Fig. \ref{fig:final_sample_selection} (respectively, the bottom-left, top-right, bottom-right, and top-left corners). For each column, galaxies are shown in decreasing order of redshift (from top to bottom).
    }\label{fig:example_cutouts}
\end{figure}

In this appendix, we present several examples of galaxy cutouts from our sample for each of the categories defined in Section \ref{sec:merger_identification} (see also Fig. \ref{sec:merger_identification}). The galaxies in each category (column) are ordered by decreasing redshift, and the F444W band is the same used for the morphological measurements described in Appendix \ref{appendix1:definition_of_morphological_parameters}.

\clearpage

\section{Tables}\label{appendix:tables}

\begin{table}[h!]
\renewcommand{\arraystretch}{1.5} 
\vspace{+0.4cm}
%\small
\caption{Merger fraction and median galaxy properties in the redshift bins analysed in this work}             
\label{table:results_merger_fractions}     
%\centering  
\vspace{-0.35cm}
\begin{center} { 
\setlength\extrarowheight{-3pt}
\begin{tabular}{ | m{0.71cm} | m{1.35cm} | m{1.01cm} | m{1.01cm} | m{1.26cm} | m{1.17cm} | } 
  \hline
  $z_{\rm med}$ & $\log_{10}$ M$_{\ast, \rm med}$/M$_\odot$ & N$_{\rm galaxies}$ & f$_{\rm m,golden}$ & f$_{\rm m,asymmetry}$ & f$_{\rm m,Gini-M_{20}}$ \\ 
  \hline
  \hline
  5.22 & 8.72 &  438 & 0.05$_{-0.01}^{+0.02}$  & 0.08$_{-0.01}^{+0.02}$  & 0.17$_{-0.02}^{+0.02}$ \\
  5.68 & 8.50 &  343 & 0.05$_{-0.01}^{+0.02}$  & 0.07$_{-0.02}^{+0.02}$  & 0.14$_{-0.02}^{+0.02}$ \\
  6.11 & 8.51 &  103 & 0.05$_{-0.02}^{+0.03}$  & 0.07$_{-0.02}^{+0.03}$  & 0.13$_{-0.03}^{+0.05}$ \\
  6.75 & 8.68 &  111 & 0.05$_{-0.02}^{+0.04}$  & 0.07$_{-0.02}^{+0.04}$  & 0.12$_{-0.03}^{+0.05}$ \\
  7.29 & 8.50 &  169 & 0.04$_{-0.02}^{+0.04}$  & 0.05$_{-0.02}^{+0.04}$  & 0.13$_{-0.03}^{+0.04}$ \\
  8.68 & 8.4  &  43  & 0.02$_{-0.01}^{+0.05}$  & 0.02$_{-0.01}^{+0.05}$  & 0.12$_{-0.05}^{+0.08}$ \\
  9.85 & 8.56 &  21  & $<0.07$               & $<0.07$               & 0.05$_{-0.04}^{+0.13}$ \\
  \hline
\end{tabular} 
}
\vspace{-0.25cm}
\tablefoot{The first three columns indicate, respectively, the median redshift, the median stellar mass, and the total number of galaxies for each redshift bin, as defined in the text. The last three columns include the merger fraction (with $1 \sigma$ uncertainties) for the three different methods adopted in this work. f$_{\rm m,golden}$ and f$_{\rm m,asymemtry}$ in the last redshift bin are only upper limits (calculated with the low number statistics of \citet{gehrels86}) as no galaxies satisfy the corresponding merger conditions. 
}
\end{center}
\vspace{-0.2cm}
\end{table}

\begin{table}[ht]
\renewcommand{\arraystretch}{1.5} 
\vspace{+0.4cm}
\small
\caption{SFR and sSFR enhancement in mergers at $z>5$}              
\label{table:results_SFR_enhancement}      
%\centering  
\vspace{-0.35cm}
\begin{center} { \small
\setlength\extrarowheight{-2pt}
\textbf{SFR for mergers and non-mergers in different stellar mass bins} 
\begin{tabular}{ | m{2.1cm} | m{1.8cm} | m{1.8cm} | m{1.8cm} | } 
  \hline
  $\log_{10}$ $M_{\ast,\rm med}$/M$_\odot$ & $\log_{10}$ SFR$_{\rm med}^{\rm NM}$ & $\log_{10}$ SFR$_{\rm med}^{M,1}$ & $\log_{10}$ SFR$_{\rm med}^{M,2}$ \\
  \hline
   & M$_\odot$/yr & M$_\odot$/yr  & M$_\odot$/yr \\
  \hline 
  7.7 &  0.04$\pm$0.04    & 0.29$\pm$0.16   & 0.2$\pm$0.3  \\
  8.1 &  0.32$\pm$0.04    & 0.49$\pm$0.09   & 0.66$\pm$0.1  \\
  8.4 &  0.58$\pm$0.04    & 0.87$\pm$0.06   & 0.77$\pm$0.1  \\
  8.6 &  0.69$\pm$0.04    & 1.01$\pm$0.19   & 1.01$\pm$0.13  \\
  8.9 &  0.9$\pm$0.05     & 1.04$\pm$0.17   & 1.15$\pm$0.09   \\
  9.4 &  1.25$\pm$0.06    & 1.23$\pm$0.23   & 1.19$\pm$0.12   \\
  \hline
\end{tabular} 
\textbf{sSFR as a function of redshift for mergers and non-mergers} 
\begin{tabular}{ | m{0.7cm} | m{1.5cm} | m{1.5cm} | m{1.75cm} | m{1.65cm} | } 
  \hline
  $z_{\rm med}$ & $\log_{10}$ sSFR$_{\rm med}^{NM,golden}$ & $\log_{10}$ sSFR$_{\rm med}^{M,golden}$ & $\log_{10}$ sSFR$_{\rm med}^{NM,Gini-M_{20}}$ & $\log_{10}$ sSFR$_{\rm med}^{M,Gini-M_{20}}$ \\
  \hline 
   & 1/Gyr & 1/Gyr & 1/Gyr & 1/Gyr \\
  \hline 
  5.1 &  1.11$\pm$0.05    & 1.4$\pm$0.18 &  0.95$\pm$0.03    & 1.28$\pm$0.11  \\
  5.3 &  1.01$\pm$0.04    & 1.13$\pm$0.17 &  0.99$\pm$0.02    & 1.1$\pm$0.11  \\
  5.6 &  1.09$\pm$0.05    & 1.4$\pm$0.2  &  1.02$\pm$0.03    & 1.24$\pm$0.15  \\
  6.0 &  1.16$\pm$0.06    & 1.4$\pm$0.16 &  1.09$\pm$0.04     & 1.05$\pm$0.12  \\
  6.8 &  1.23$\pm$0.04     & 1.45$\pm$0.22 &  1.21$\pm$0.03    & 1.37$\pm$0.13   \\
  7.6 &  1.4$\pm$0.04    & 1.42$\pm$0.19 &  1.36$\pm$0.02    & 1.43$\pm$0.11   \\
  \hline
\end{tabular} 
}
\vspace{-0.25cm}
\tablefoot{The upper table includes the median SFR in M$_\ast$ bins for the golden merger sample (label $^{M,golden}$), Gini-M$_{20}$ mergers (label $^{M,Gini-M_{20}}$), and non-merger sample (label $^{NM}$), as shown in Fig.\ref{fig:SFR_mass}. The lower table includes the median sSFR in $z$ bins for the merger and non-merger samples, as shown in Fig.\ref{fig:SSFR_redshift}. The labels $^{NM,golden}$ and $^{NM,Gini-M_{20}}$ refer to the control sample of non-merger galaxies of the golden merger sample and Gini-M$_{20}$ mergers, respectively. 
}
\end{center}
\vspace{-0.2cm}
\end{table}

\begin{table}[ht]
\renewcommand{\arraystretch}{1.5} 
\vspace{+0.4cm}
%\small
\caption{Merger rates and specific mass accretion rates due to mergers for the sample analyzed in this work}  
\label{table:results_merger_rates_sMAR}     
%\centering  
\vspace{-0.35cm}
\begin{center} { 
\setlength\extrarowheight{-3pt}
\begin{tabular}{ | m{1.2cm} | m{1.3cm} | m{1.3cm} | m{1.3cm} | m{1.3cm} | } 
  \hline
  $z_{\rm med}$ & $\mathcal{R}_\mathrm{M}^1$ & sMAR$^1$ & $\mathcal{R}_\mathrm{M}^2$ & sMAR$^2$ \\ 
  \hline
  \hline
  5.22 &  1.6$_{-0.4}^{+0.1}$    & 0.8$_{-0.4}^{+1.2}$ &  4.70$_{-0.15}^{+1.8}$ & 0.7$_{-0.2}^{+0.8}$  \\
  5.77 &  1.9$_{-0.5}^{+0.2}$    & 0.9$_{-0.5}^{+1.4}$ &  4.62$_{-0.15}^{+1.8}$ & 0.7$_{-0.2}^{+0.8}$  \\
  6.20 &  1.8$_{-0.4}^{+0.2}$    & 0.9$_{-0.5}^{+1.4}$ &  4.54$_{-0.15}^{+1.8}$ & 0.7$_{-0.2}^{+0.8}$  \\
  6.73 &  2.2$_{-0.5}^{+0.2}$    & 1.1$_{-0.6}^{+1.6}$ &  4.76$_{-0.16}^{+1.9}$ & 0.7$_{-0.2}^{+0.8}$  \\
  7.28 &  2.2$_{-0.5}^{+0.2}$    & 1.1$_{-0.6}^{+1.6}$ &  6.00$_{-0.19}^{+2.3}$ & 0.9$_{-0.3}^{+1}$    \\
  8.50 &  1.0$_{-0.2}^{+0.08}$   & 0.5$_{-0.3}^{+0.8}$ &  6.5$_{-0.2}^{+2.5}$   & 1.0$_{-0.3}^{+1}$    \\
  9.71 &  $<5.0$               & $<4.3$            &  3.23$_{-0.11}^{+1.3}$ & 0.5$_{-0.2}^{+0.5}$  \\
  \hline
\end{tabular} 
}
\vspace{-0.25cm}
\tablefoot{The first column is the median redshift of each bin. The second and fourth columns represent the merger rate. The third and fifth ones are the specific mass accretion rate due to mergers. The label $^1$ refers to major mergers (our gold merger sample), while the label $^2$ refers to the total mergers satisfying the condition Gini $+ 0.14 \times $ M$_{20}$ $> 0.33$. 
}
\end{center}
\vspace{-0.2cm}
\end{table}

\begin{table*}[t!]
\renewcommand{\arraystretch}{1.5} 
\vspace{+0.4cm}
%\small
\caption{Comparison studies, grouped according to the merger identification method}                 
\label{table:comparison_works}      
\vspace{-0.35cm}
\begin{center} { %\small
\setlength\extrarowheight{-3pt}
\begin{tabular}{ | m{3.6cm} | m{2.4cm} | m{1.2cm} | m{2.5cm} | m{6.6cm} | } 
  \hline
  Study & imaging data & redshift range  & Mass range [$\log$ M$_{\ast}$/M$_\odot$] & merger selection \\ 
  \hline
  \hline
  Ellison et al. 2013 & SDSS & 0.01-0.2 & [9-12] & visual morphology \\
  Kaviraj et al. 2013 & HST/WFC3 & 1.9-2.1 & $>10$ & visual morphology \\ % (UVIS + IR bands)
  Robaina et al. 2009 & HST/ACS & 0.4-0.8 & $>10$ & visual morphological disturbances + pairs (d$_{\rm sep}<$ [$15$ kpc) \\
  \hline
  Kartaltepe et al. 2007 & HST + ground based telescopes & 0 - 1.2 & M$_V < -19.8$ & close pairs with $\delta(z)<0.05$ and d$_{\rm sep}=$ [$5$-$20$] kpc \\
  Patton et al. 2013 & SDSS & 0.02-0.2 & [9-12] (median $\sim$ 10.7) & galaxy pairs \\
  Man et al. 2016 & HST & $0.1$ - 3 & $>10.8$ &  close pairs with $\delta(z)/(1+z)<0.1$ and d$_{\rm sep}=$ [$10$-$30$] $h^{-1}$ kpc \\
  Mantha et al. 2018 & HST  & 0 - 3 & $>10.3$ &  close pairs with d$_{\rm sep}=$ [$5$-$50$] kpc and redshift proximity criterion \\
  Duan et al. 2025 & JWST/NIRCam & 4.5-11.5 & [8-10] & close pairs (probabilitistic method based on P(z)) with d$_{\rm sep}$ = [$20$-$50$] kpc \\
  Puskas et al. 2025 & JWST/NIRCam & 3 - 9 & [8-10] & close pairs with d$_{\rm sep}=$ [$5$-$30$] kpc and $\delta(z)/(1+z)<0.1$ \\
  \hline
  Conselice et al. 2003 & HST (rest-frame B bands) & 0 - 3 &  $>8$ (M$_B$ $<-18$) & $A>0.35$ \\
  Conselice et al. 2009 & HST (F814W) & 0.2 - 1.2 & $>10$ & $A>0.35$ and $A>S$ \\
  Lopez-Sanjuan +2009 & HST (rest-frame B bands) & 0 - 1 & >$10$ (M$_B$<$-19.5$) & $A>0.35$ \\
  Conselice \& Arnold 2009 &  HST (z-band) & 3.5 - 6 & $> 9$-$10$ & $A>0.35$ \\
  \hline
  Lotz et al. 2008 & HST (F606W + F814W) & 0.2-1.2 & L$_B > 0.4 \times$ L$_B^\star$ (approx. M$_B$ $<-20.5$ & Gini$+0.14 \times M_{20} > 0.33$ \\ % and S/N$_{\rm pixel} > 2.5$ \\
  Jogee et al. 2009 & HST (F606W to F850LP) & 0.24-0.8 & $>10.4$ & quantitative + visual classification (selection of major+minor mergers) \\
  \hline
  Dalmasso et al. 2024  & JWST/NIRcam (F150W+F200W) & 4 - 9 & [7-10] (median:$8.5$) & Gini$+0.14 \times M_{20} > 0.33$ and $A > 0.35$ \\
  \hline
\end{tabular} 
}
\vspace{-0.25cm}
\tablefoot{The first two columns indicate the study and the imaging data adopted for the merger analysis. The second and third columns indicate the redshift range analyzed and the stellar mass (or absolute magnitude) range of the analyzed sample. The last columns specifies the criterion adopted for merger classification. d$_{\rm sep}$ is the projected separation distance between galaxies for the photometric close pair method. 
}
\end{center}
\vspace{-0.3cm}
\end{table*}

\end{document}